\documentclass[12pt,preprint]{elsarticle} 
\usepackage{bm} 
\usepackage{amsmath,amssymb}
\usepackage{mathrsfs}
\usepackage{amsbsy}
\usepackage{color}
\usepackage[english]{babel}
\usepackage{caption}
\usepackage{subcaption}
\usepackage{natbib}
\usepackage{tikz}
\usepackage{subfig}
\usetikzlibrary{arrows}
\usepackage{pstricks-add}
\usepackage{graphicx}
\usepackage{epstopdf}
\usepackage[toc,page]{appendix}
\biboptions{sort&compress,authoryear}

\newcommand{\br}{\mathbf{r}}

\newcommand{\bu}{\mathbf{u}}

\newcommand{\bw}{\mathbf{w}}

\newcommand{\bL}{\mathbf{L}}

\newcommand{\bT}{\mathbf{T}}

\newcommand{\bzero}{\mathbf{0}}

\DeclareMathOperator{\sinc}{sinc}

\begin{document}
\begin{frontmatter}

\title{Non-Reciprocal Wave Propagation in Spatiotemporal Periodic Structures}

\author[DG]{G. Trainiti\corref{cor1}}

\author[DG,MS]{M. Ruzzene}

\address[DG]{Daniel Guggenheim School of Aerospace Engineering, Georgia Institute of Technology, 270 Ferst Dr, Atlanta, Georgia 30332, USA}
\address[MS]{George W. Woodruff School of Mechanical Engineering, Georgia Institute of Technology, 801 Ferst Dr, Atlanta, Georgia 30332, USA}
\cortext[cor1]{Corresponding author at: Daniel Guggenheim School of Aerospace Engineering, Georgia Institute of Technology, 270 Ferst Dr, Atlanta, Georgia 30332, USA. Email: gtrainiti@gatech.edu - Tel.:+1 404 786 8489}

\begin{abstract}
We study longitudinal and transverse wave propagation in beams with elastic properties that are periodically varying in space and time. Spatiotemporal modulation of the elastic properties breaks mechanical reciprocity and induces one-way propagation. We follow an analytic approach to characterize the non-reciprocal behavior of the structures by analyzing the symmetry breaking of the dispersion spectrum, which results in the formation of directional band gaps and produces shifts of the First Brilloin Zone limits. This approach allows us to relate position and width of the directional band gaps to the modulation parameters. Moreover, we identify the critical values of the modulation speed to maximize the non-reciprocal effect. We numerically verify the theoretical predictions by using a finite element model of the modulated beams to compute the transient response of the structure. We compute the two-dimensional Fourier transform of the collected displacement fields to calculate numerical band diagrams, showing excellent agreement between theoretical and numerical dispersion diagrams.
\end{abstract}
\end{frontmatter}

\newpage
\section{Introduction}
Reciprocity of wave propagation is a fundamental principle of many wave phenomena and applies in electromagnetism~\cite{Landau}, optics \cite{Helmholtz} and acoustics\cite{Pierce}, to name a few. Loosely, it states that waves propagate symmetrically in space from one point to the other, no matter which one is the source or the receiver. A growing area of research is concerned with the possibility of breaking this form of symmetry in order to realize one-way propagation which is highly desirable in many technological applications. In acoustics, for instance, it might be expedient to protect a source from its echo or achieve full-duplex sound communication~\cite{NonReciprocalAcoustics}. Furthermore, non-reciprocal devices can be used to realize one-way filters and isolation \cite{Cummer2016,NonReciprocalAcoustics}. In mechanical systems, weak non-reciprocal wave propagation occurs in rotating rings, in which the rotation introduces a mechanical bias that leads to different wavenumber and phase velocity values for waves propagating in opposite directions~\cite{Huang20134979}. The mechanical bias is due to Coriolis forces~\cite{Beli2015}, which are responsible for breaking reciprocity~\cite{PhysRev.37.405}. A similar non-reciprocal behavior is observed in systems subjected to a magnetic field~\cite{PhysRev.38.2265}. Strong non-reciprocal effect are achieved by relaxing some of the assumptions of the Onsager-Casimir principle of microscopic reversibility \cite{RevModPhys.17.343}, which states that reciprocity is not guaranteed when nonlinearity or time-dependent material properties are exploited~\cite{NonReciprocalAcoustics}. Therefore, considerable efforts have been devoted to achieve non-reciprocal behavior through nonlinear or time-modulated devices. Indeed, giant non-reciprocal transmission can be obtained by combining a nonlinear medium with a superlattice\cite{Liang2009,Liang2010} or with a gain/loss pair~\cite{Liang2016}, hence exploiting asymmetric frequency conversion of the nonlinear medium due to second-harmonic generation (SHG) and frequency selectivity of sonic crystals. Experimental evidence shows rectifying ratios up to $10^4$, although efficacy of such nonlinear devices depends on the amplitude of the input signal, which has to be large enough to trigger the SHG mechanism of the nonlinear medium~\cite{Liang2010}. Another way to exploit nonlinearity is to use a compact active acoustic metamaterial coupled to a nonlinear electronic circuit~\cite{Popa2014}, which produces an isolation factor $> 10$ dB. Several recent studies have investigated the possibility of modifying in time the material properties of the system to achieve non-reciprocal behavior. For example, one-way acoustic isolation has been shown in graphene based nanoelectromechanical systems (NEMS)\cite{Zanjani2015}, acoustic waveguides~\cite{Zanjani2014}, acoustic circulators~\cite{fleury2014sound} and time-dependent superlattices~\cite{Swinteck2015}. The common strategy in these studies is to extend to the time domain the spatial-only periodic variation of material properties that in general is associated with metamaterials. This strategy thus exploits spatiotemporal modulated systems to manipulate wave propagation. Space and time modulation of the medium supporting wave propagation already gained the attention of the scientific community many decades ago. For instance, Slater~\cite{Slater} studied the scattering of an electronic wave by a sinusoidal perturbation. In this problem, a particle is moving in a region with potential energy having the form of a traveling plane wave $V=V_1 \cos (\omega_1 t - \bm{\kappa}_1 \cdot \br)$. The Schr{\"o}edinger's equation for this particle, in the one dimensional case, reduces to a Mathieu's equation~\cite{brillouin1953wave}. A perturbation approach is employed to obtain the first three harmonics $n=0,\pm1$ of the wave function, under the assumption that the amplitude of modulation $V_1$ of the potential energy is small. Moreover, an approximate relation between the energy $\hbar\omega_0$ and the wavevector $\bm{\kappa}_0$ is derived, which is able to predict the first order stop-band or Bragg reflection, but fails at describing higher-order stop-bands. Simon~\cite{Simon} and Hessel et al.~\cite{Hessel1961} suggested a more general approach to study the wave propagation of an electromagnetic wave in a medium with sinusoidal disturbance, i.e. with dielectric constant $\epsilon=\epsilon_0 + \epsilon_1 \cos(\omega_1 t - \bm{\kappa}_1 \cdot \br)$. The wave equation for the electric field is solved by imposing a Floquet solution with space-time harmonics, which leads to a description of higher-order stop-bands by computing the dispersion relation given in the form of a rapidly convergent continued fraction. While many of these and others pioneering works focused mainly on issues related to traveling wave parametric amplification~\cite{Cullen1958,Tien1958} and the condition for stability of the system associated to frequency conversion effects \cite{Cassedy1963,Cassedy1967}, to the best of the authors' knowledge, the possibility of achieving one-way propagation in spatiotemporal modulated system was not properly recognized and described.
In the present work, we describe the dispersion diagrams of elastic waves propagating in beams with space-time periodic material properties. We study both non-dispersive longitudinal motion and dispersive transverse motion. The dispersion diagrams are used to identify a new class of band gaps that allow wave propagation in one direction only, thus they are referred to as non-reciprocal or directional band gaps. In one-dimensional systems, the directional band gaps signal and quantify the non-reciprocal behavior of the structure. We show that reciprocity is broken when a spatiotemporal modulation of the material properties of the structure, \emph{i.e.} the Young's modulus $E$ and $\rho$, is imposed in the form of a traveling wave. In this case, we assume a solution in the Floquet form with space-time harmonics, following the approach of \cite{Slater}. The Floquet solution is substituted into the general equations of motion of beams with space and time varying material properties and leads to a Quadratic Eigenvalues Problem\cite{QEP} (QEP) that can be solved for the frequency $\omega$ as a function of the wavenumber $\kappa$. This approach also allows us to obtain approximate analytical relationships between the modulation parameters and the position and width of the directional band gaps for the simple case of harmonic modulation. Moreover, we compute the minimum speed at which the modulating traveling wave has to travel in order to maximize the non-reciprocal effect. We also discuss the shift of the First Brilloin Zone limits due to spatiotemporal modulation.
The paper is divided into 4 sections. The present introduction (Section 1) is followed by Section 2, which firstly gives a description of the systems under investigation, the relative equations of motion and the considered modulation strategies, secondly expounds the methodology employed for the analysis. Results are presented in Section 3 for both longitudinal and transverse motion. Concluding remarks are presented in Section 4.
\clearpage

\section{Theoretical background}\label{TheorySection}
\subsection{Time-spatial periodic beams}\label{subsec:TimeSpatialPS}
Consider a beam with Young's modulus $E$ and density $\rho$ periodic functions of space and time.
We define this periodic variation of the medium's characteristic in space and time as modulation. We consider a periodic variation along the axial direction, thus the modulated Young's modulus $E$ and density $\rho$ are expressed as:
\begin{equation}
E(x,t) = E(x + \lambda_m,t + T_m) \qquad \qquad \rho(x,t) = \rho(x + \lambda_m,t + T_m) 
\end{equation}
for any location $x$ and at any instant $t$, in which $\lambda_m$ and $T_m$ define the spatial and time periodicity, respectively, while $\kappa_m=2\pi/\lambda_m$ and $\omega_m=2\pi/T_m$ are the wavenumber and the angular frequency associated to the properties of modulation. The periodic modulation pattern can be understood as a traveling wave with velocity $v_m=\omega_m/\kappa_m$. We limit our analysis to the case in which this pattern travels with constant velocity and we define $v_m$ as ``modulation speed''.
\begin{figure}[hbtp]
	\centering          
	\begin{subfigure}[b]{0.48\textwidth}
		\includegraphics[width=\textwidth]{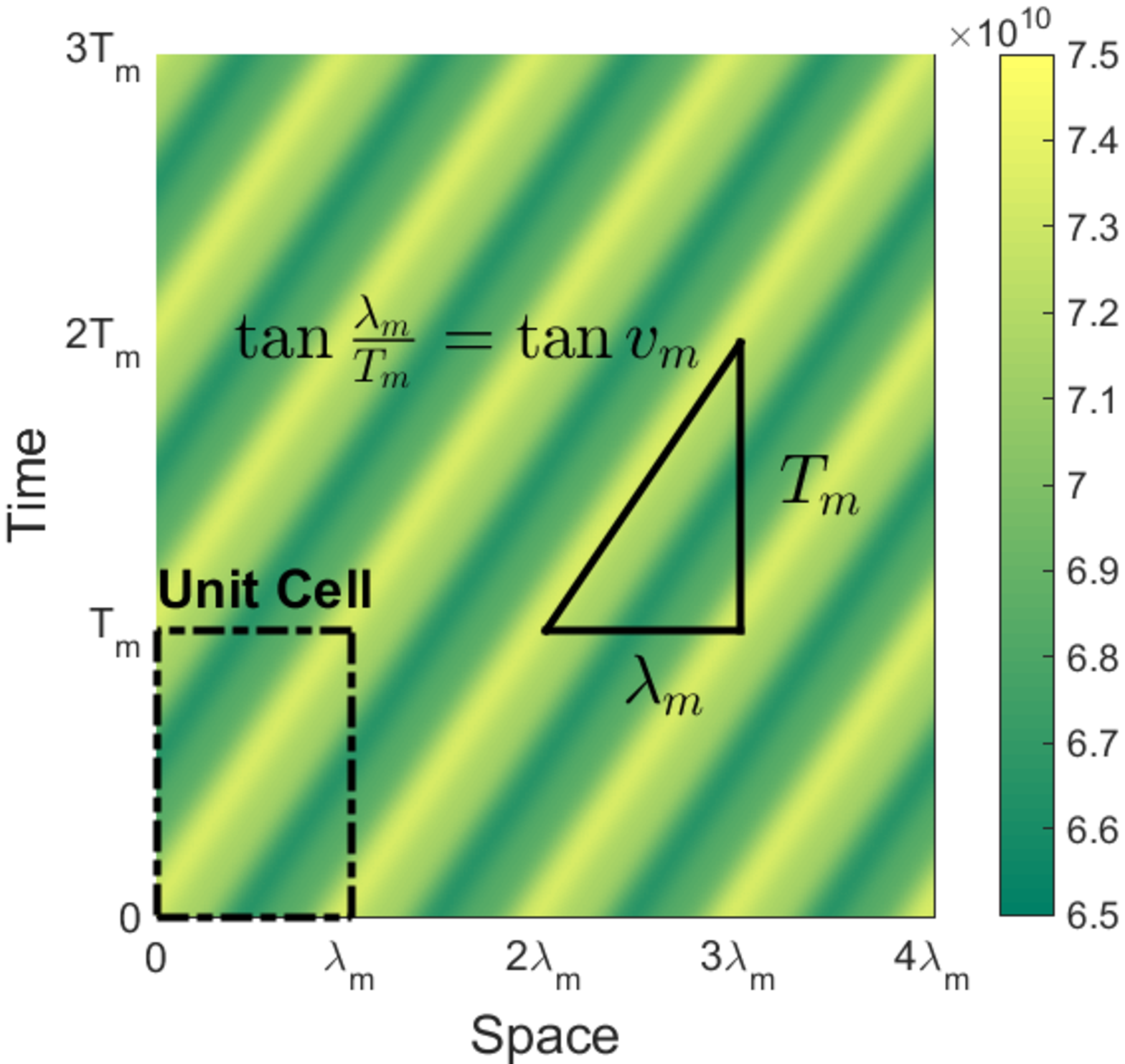}
		\caption{}
		\label{Fig:General_ST_UC}
	\end{subfigure}
	\begin{subfigure}[b]{0.48\textwidth}
		\includegraphics[width=\textwidth]{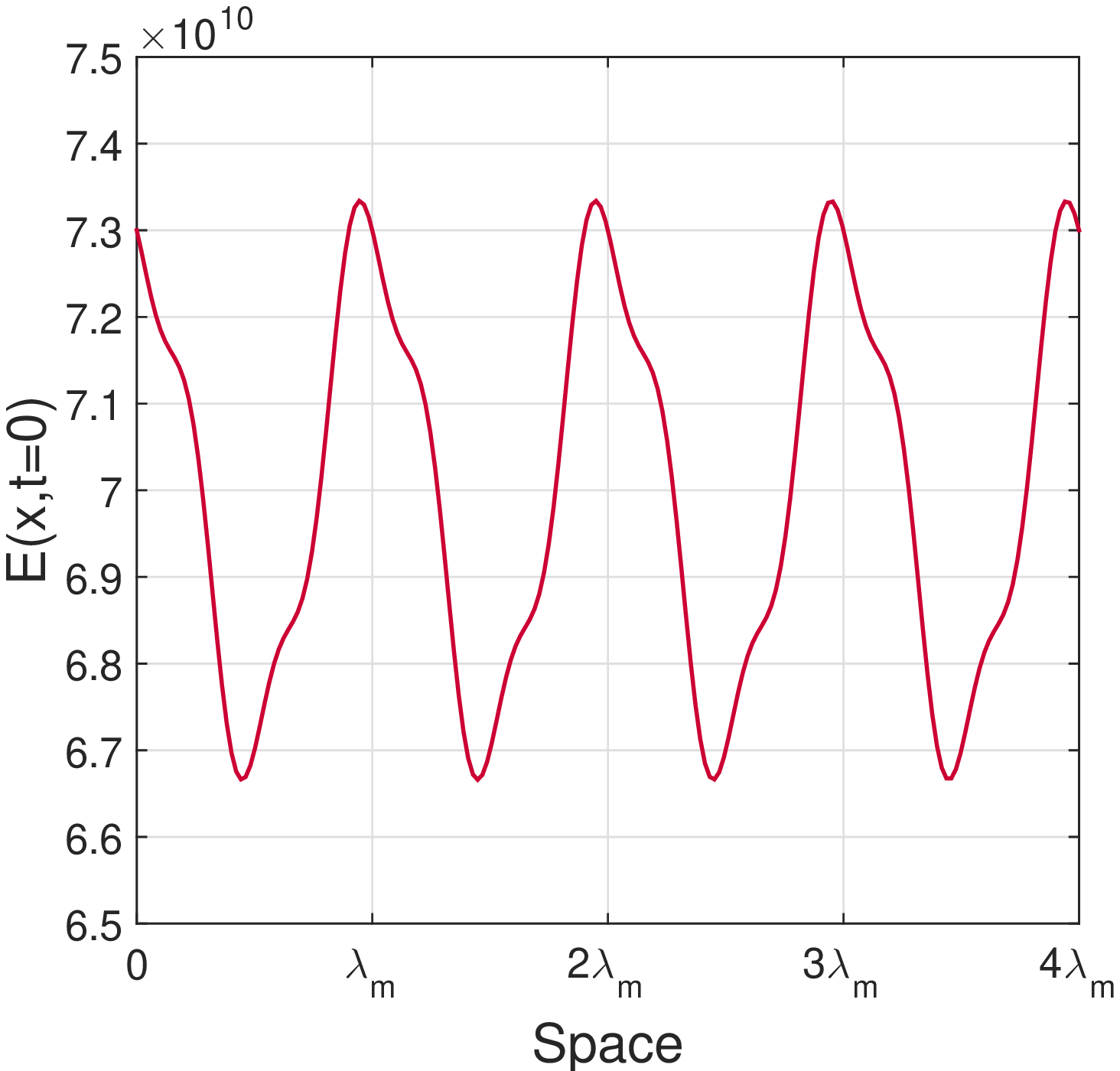}
		\caption{}
		\label{Fig:General_ST_UC_Snapshot}
	\end{subfigure}               
	\caption{Example of spatiotemporal periodic material with $E(x,t)=E_0 + E_m\cos(\omega_m t - \kappa_m x) + E_m/5\cos\big[3(\omega_m t - \kappa_m x)\big]$, with $E_0=70$ GPa and $E_m=3$ GPa: (a) spatiotemporal unit cell and its periodic space-time domain; (b) profile of the traveling modulation pattern as see at $t=0$.}
	\label{Fig:General_ST_UC_Tot}
\end{figure}
The behavior of spatial-only periodic structures can be fully characterized by studying a single unit cell, which represents the building block of the structure~\cite{brillouin1953wave,kittel2004introduction}. For systems periodic both in space and time, the concept of unit cell has to be extended to account for periodicity both in the spatial and temporal domains. Therefore, in our study we focus on the spatiotemporal unit cell in order to obtain the dispersion properties of the structure. The spatiotemporal unit cell is associated to $\lambda_m$ and $T_m$, it travels with velocity $v_m$ along the axial direction and coincides with the classically defined unit cell of spatial-only periodic structures for $v_m=0$. An example of spatiotemporal unit cell is given in Fig.~\ref{Fig:General_ST_UC}, in which we consider a material with Young's modulus given by $E(x,t)=E_0 + E_m\cos(\omega_m t - \kappa_m x) + E_m/5\cos\big[3(\omega_m t - \kappa_m x)\big]$, with $E_0=70$ GPa and $E_m=3$ GPa. At $t=0$, the stiffness of the structure is defined by the periodic function shown in Fig.~\ref{Fig:General_ST_UC_Snapshot}. We remark here that the unit cell can travel in both forward and backward directions. In the present study, we consider forward propagating modulation only, since analogous considerations can be made for backward propagating modulation.
\subsection{Analysis of dispersion}
We first derive the equations of motions of beams with material properties that depend upon space and time. We consider both longitudinal and transverse motion. The study is restricted to the case of slender beams, in which shear deformation and rotational inertia are neglected~\cite{Meirovitch}. Time-space periodic structures are then investigated by restricting the class of functions that describe the material properties to periodic functions in time and space only.
\begin{figure}
  \centering
    \includegraphics[width=0.48\textwidth]{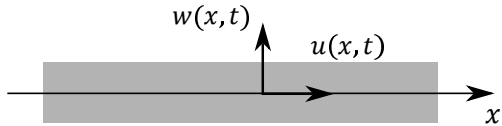}
      \caption{Schemtic of a beam with longitudinal motion described by $u(x,t)$ and transverse motion described by $w(x,t)$.}
      \label{fig:BeamSchematic_Theory}
\end{figure}
As shown in Fig.~\ref{fig:BeamSchematic_Theory}, in the case of longitudinal motion we assume $u=u(x,t)$ to be the longitudinal displacement of the beam along the axial direction. For transverse motion, the displacement is described by $w=w(x,t)$. We impose the conservation of linear momentum along both axial and transverse direction and the constitute laws for the cross-section of the beam to get the following equation of motion for longitudinal motion:
\begin{equation}
\frac{\partial}{\partial x}\bigg[E(x,t) \frac{\partial u (x,t)}{\partial x}\bigg] - \frac{\partial}{\partial t}\bigg[\rho(x,t) \frac{\partial u (x,t)}{\partial t}\bigg] = 0 \label{eq:eqmot2}
\end{equation}
and transverse motion:
\begin{equation}
R^2_g\frac{\partial^2}{\partial x^2}\bigg[E(x,t)\frac{\partial^2 w(x,t)}{\partial x^2}\bigg]+\frac{\partial}{\partial t}\bigg[\rho(x,t) \frac{\partial w (x,t)}{\partial t}\bigg]=0 \label{eq:eqmot2Beam}
\end{equation}
in which $R^2_g=I/A$ is the radius of gyration of the beam. If $E(x,t)=E$ and $\rho(x,t)=\rho$ are constant in space and time, Eq.~\eqref{eq:eqmot2} and Eq.~\eqref{eq:eqmot2Beam} reduce to the familiar equations of motion for uniform beams.

We study wave propagation properties in time-spatial periodic structures by computing and analyzing band diagrams for said structures. The approach followed in this work exploits the periodic nature of the considered modulation of the material properties functions $E(x,t)$ and $\rho(x,t)$. Both are periodic in the variables $x$ and $t$ with periodicity $\lambda_m$ and $T_m$, respectively. It is then possible to express them by using the following Fourier series representations:
\begin{equation}\label{eq:permatseries}
E(x,t) =  \sum\limits_{p=-\infty}^{+\infty} \hat{E}_p e^{i p (\omega_m t - \kappa_m x)} \qquad \qquad \rho(x,t) =  \sum\limits_{p=-\infty}^{+\infty} \hat{\rho}_p e^{i p (\omega_m t - \kappa_m x)}
\end{equation}
in which $\hat{E}_p$ and $\hat{\rho}_p$ are the Fourier coefficients of the series associated with the traveling harmonic terms $e^{i p (\omega_m t - \kappa_m x)}$. An explicit expression for these coefficients can be obtained by performing a double integration over the unit cell's spatial and time periods $\lambda_m$ and $T_m$, respectively:
\begin{align}\label{Eq:FourierCoeff}
\hat{E}_p &= \frac{1}{T_m}\frac{1}{\lambda_m}\int_{-T_m/2}^{+T_m/2}\int_{-\lambda_m/2}^{+\lambda_m/2} E(x,t) e^{-ip(\omega_m t - \kappa_m x)} dxdt
\\
\hat{\rho}_p &= \frac{1}{T_m}\frac{1}{\lambda_m}\int_{-T_m/2}^{+T_m/2}\int_{-\lambda_m/2}^{+\lambda_m/2} \rho(x,t) e^{-ip(\omega_m t - \kappa_m x)} dxdt
\end{align}
The solutions to Eq.~\eqref{eq:eqmot2} and Eq.~\eqref{eq:eqmot2Beam} are chosen in the generalized Floquet form~\cite{Slater} and write:
\begin{align}
u(x,t)=&e^{i(\omega t - \kappa x)} \sum\limits_{n=-\infty}^{+\infty} \hat{u}_n e^{in(\omega_m t - \kappa_m x )}\label{eq:sol1}\\
w(x,t)=&e^{i(\omega t - \kappa x)} \sum\limits_{n=-\infty}^{+\infty} \hat{w}_n e^{in(\omega_m t - \kappa_m x )}\label{eq:sol2}
\end{align}
In a study on wave propagation in spatiotemporal periodic transmission lines~\cite{Cassedy1967}, allowing solutions in the generalized Floquet form, it is discussed the possibility of having time-growing waves for certain modulation parameters. These time-growing waves are associated to unstable interactions between the wave propagating in the system and the modulation wave. In the same study, the author discusses under which conditions the governing equations of the system are satisfied, for real wavenumbers, by complex frequencies only, leading to time-growing oscillations. Such instabilities occur when the modulation speed is greater than the wave speed in the uniform system. Although interesting, an analogous study of the stability conditions for wave propagation in modulated beams falls outside the scope of the present article. Nevertheless, knowing that non-periodic, time-growing solutions can arise for certain combinations of the modulation parameters $\omega_m$, $\kappa_m$ and $\hat{E}_p$, $\hat{\rho}_p$, we assume stability to hold by imposing a modulation speed smaller than the wave speed in the uniform structure.
Substituting Eq.~\eqref{eq:sol1} and Eq.~\eqref{eq:sol2} into Eq.~\eqref{eq:eqmot2} and Eq.~\eqref{eq:eqmot2Beam}, together with the expressions for the material properties in Eq.~\eqref{eq:permatseries}, leads to expressions that can be simplified by exploiting orthogonality of the Fourier basis in Eq.s~\eqref{eq:eqmot2}-\eqref{eq:eqmot2Beam}. This gives:
\begin{align}
\sum\limits_{n=-\infty}^{+\infty}& \bigg[ \big( \kappa + n \kappa_m\big) \big( \kappa + p \kappa_m\big) \bigg] \hat{E}_{p-n} \hat{u}_n= \sum\limits_{n=-\infty}^{+\infty} \bigg[ \big( \omega + n \omega_m\big) \big( \omega + p \omega_m\big) \bigg] \hat{\rho}_{p-n} \hat{u}_n\label{eq:rawQEP1}\\
\sum\limits_{n=-\infty}^{+\infty}& \bigg[ \big( \kappa + n \kappa_m\big) \big( \kappa + p \kappa_m\big) \bigg]^2 \hat{E}_{p-n} \hat{w}_n= \frac{1}{R^2_g} \sum\limits_{n=-\infty}^{+\infty} \bigg[ \big( \omega + n \omega_m\big) \big( \omega + p \omega_m\big) \bigg] \hat{\rho}_{p-n} \hat{w}_n\label{eq:rawQEP2}
\end{align}
which hold for the longitudinal and transverse motion of the beam, respectively. If a finite number $N$ of terms is considered in Eq.~\eqref{eq:sol1} and~\eqref{eq:sol2}, then Eq.~\eqref{eq:rawQEP1} and~\eqref{eq:rawQEP2} represent each a finite set of $2N+1$ equations that can be cast in the form of a quadratic eigenvalue problem (QEP):
\begin{align}\label{eq:QEP_L}
\big(\omega^2 \hat{\bL}_2 + \omega \hat{\bL}_1 + \hat{\bL}_0 \big) \hat{\bu} &= \bzero \\  \big(\omega^2 \hat{\bT}_2 + \omega \hat{\bT}_1 + \hat{\bT}_0 \big) \hat{\bw} &= \bzero \label{eq:QEP_T}
\end{align}
where $\hat{\bL}_j=\hat{\bL}_j(\kappa)$ and $\hat{\bT}_j=\hat{\bT}_j(\kappa)$, with $j=0,1,2$, are coefficient matrices whose entries depend on the modulation parameters $\omega_m$ and $\kappa_m$, the Fourier coefficients $\hat{E}_p$ and $\hat{\rho}_p$ and the wavenumber $\kappa$. Moreover, $\hat{\bu}$ and $\hat{\bw}$ are the displacement coefficient vectors. The QEP in Eq.s~\eqref{eq:QEP_L}-\eqref{eq:QEP_T} is solved in terms of $\omega$ by letting $\kappa$ vary in a convenient range. For spatial-only periodic structure, this range is given by the Irriducible Brillouin Zone (IBZ)~\cite{brillouin1953wave}. We show that spatiotemporal modulation challenges the classical definition of the IBZ, leading to a shift in the wavenumber range to be considered. The dispersion properties of the time-spatial periodic structure are then obtained by representing the relation $\omega=\omega(\kappa)$ in the form of band diagrams.
\section{Results}\label{ResultsSection}
The results are presented in the form of band diagrams for both longitudinal and transverse wave. We also numerically verify the proposed dispersion analysis through a finite element study of the transient response to harmonic excitation of beams with modulated properties. We consider two distinct modulation strategies of the Young's modulus $E$ only, \emph{i.e.} harmonic and square traveling wave, and we assume the density $\rho$ to be constant. In discussing the results, we refer to the dimensionless frequency $\Omega$ and wavenumber $\mu$:
\begin{equation}
\Omega = \frac{f \lambda_m}{c_0} \qquad \mu=\kappa \lambda_m
\end{equation}
where $f$ is the frequency and $c_0=\sqrt{E_0/\rho_0}$ is the velocity of longitudinal waves in a non-modulated beam.

We first consider a harmonic modulation of the Young's modulus only, thus assuming the following expressions for the material parameters:
\begin{equation}\label{Eq:HM}
E(x,t) = E_0 + E_m \cos(\omega_m t - \kappa_m x), \quad\quad\quad \rho(x,t) = \rho_0,
\end{equation}
where $E_m$ is the amplitude of the modulation.
\begin{figure}[hbtp]
	\centering          
	\begin{subfigure}[b]{0.48\textwidth}
		\includegraphics[width=\textwidth]{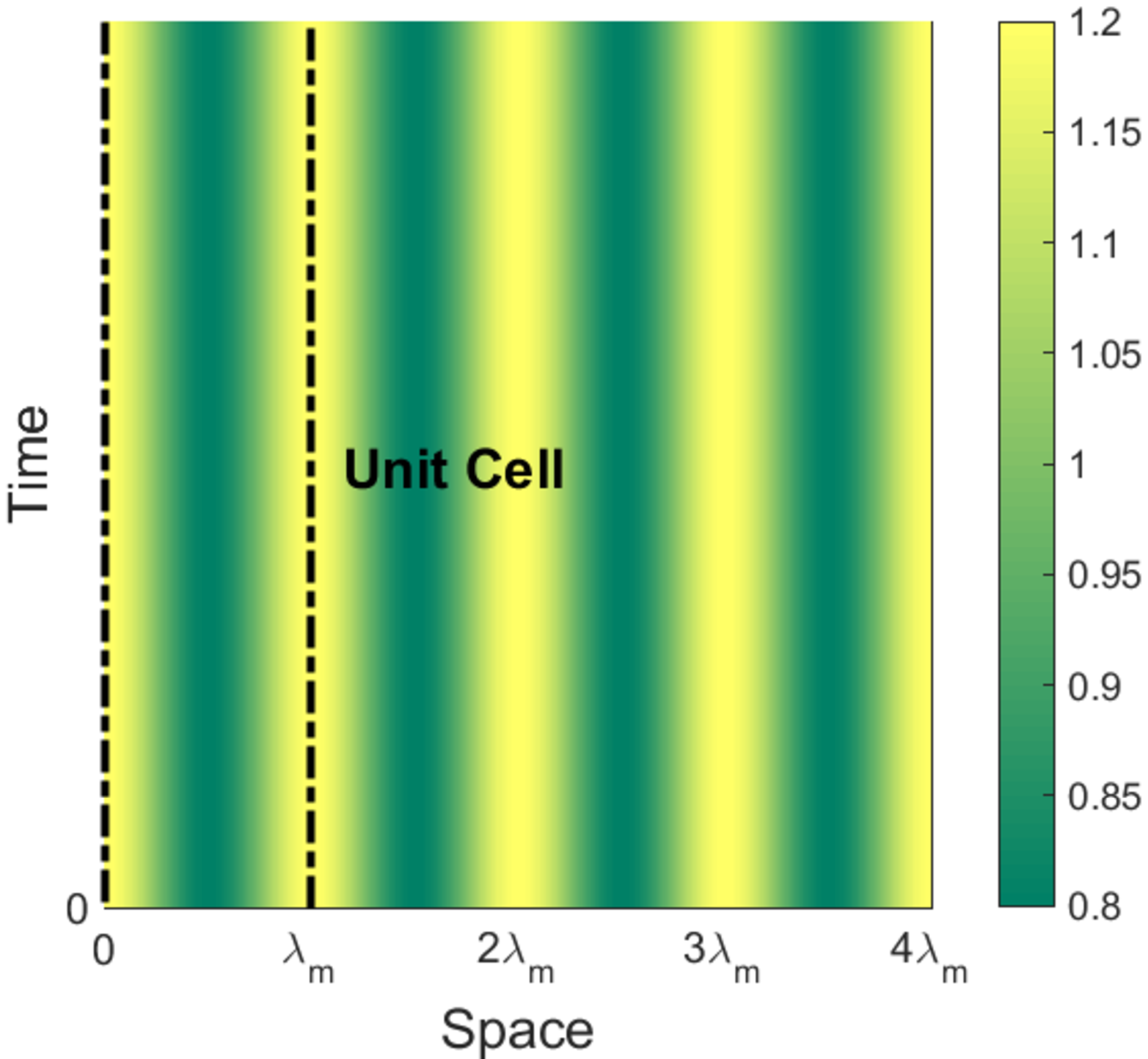}
		\caption{}
		\label{Fig:H_S_UC}
	\end{subfigure}
	\begin{subfigure}[b]{0.48\textwidth}
		\includegraphics[width=\textwidth]{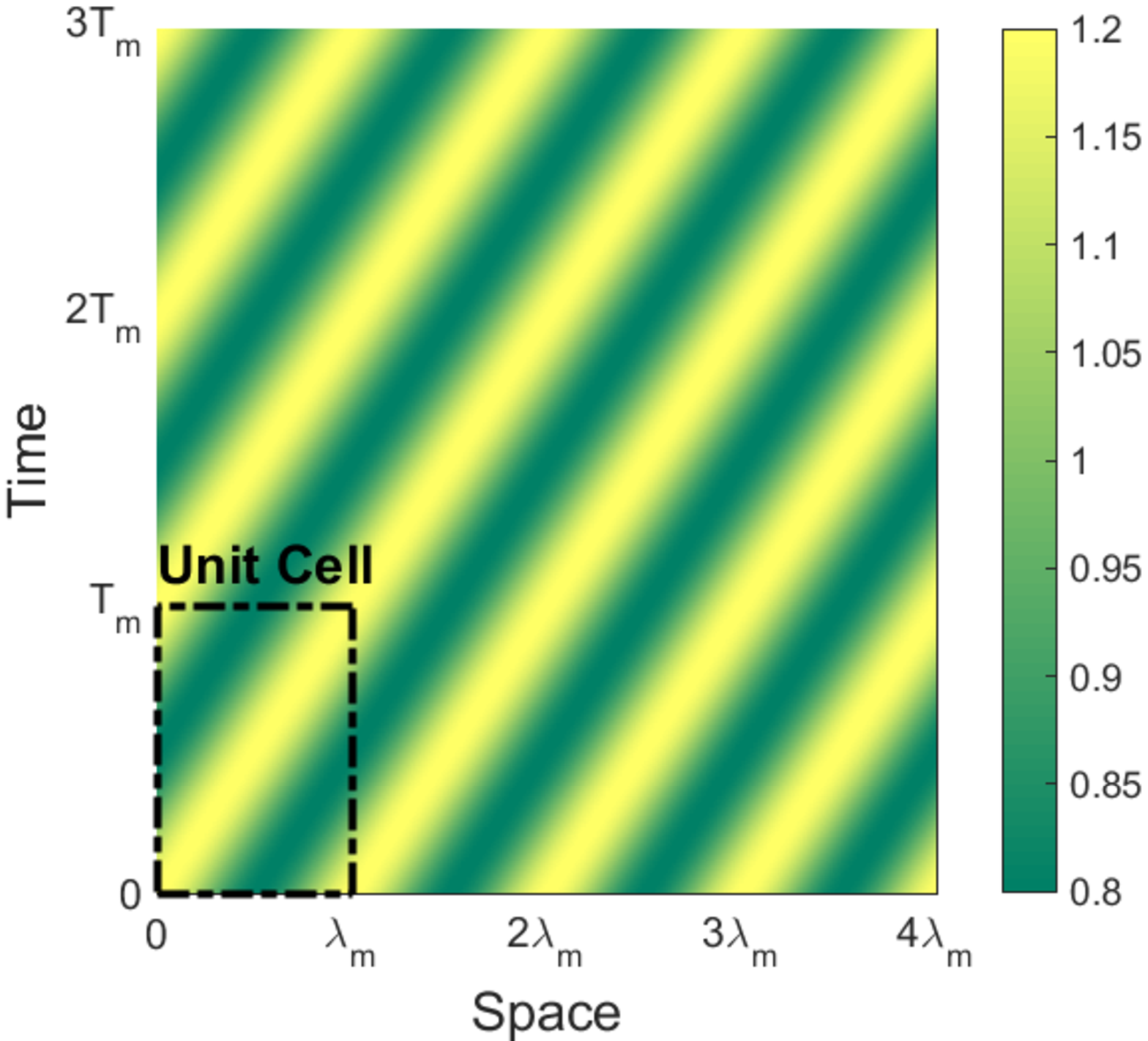}
		\caption{}
		\label{Fig:H_ST_UC}
	\end{subfigure}               
	\caption{For harmonic modulation, unit cells in the space-time domain given by the periodic change of the ratio $E(x,t)/E_0$, with $\alpha_m=0.20$: (a) spatial-only modulation ($v_m=0$, $T_m\rightarrow\infty$); (b) spatiotemporal modulation ($v_m\neq0$).}
	\label{Fig:BeamSchematicHarmonicTot}
\end{figure}
For $E_m=0$, the material is not modulated and the beam is uniform, while for $E_m\neq0$ and $\omega_m\neq0$, the value of the Young's modulus can be described as the harmonic traveling wave defined in Eq.~\eqref{Eq:HM}. The wave travels along the positive direction of the $x$ axis with speed $v_m=\omega_m/\kappa_m$.
We define the dimensionless modulation amplitude $\alpha_m$ and the dimensionless modulation velocity parameter $\nu_m$:
\begin{equation}
\alpha_m = \frac{E_m}{E_0}, \qquad   \nu_m = \frac{v_m}{c_0}.
\end{equation}
An example of harmonic modulation is given in Fig.~\ref{Fig:BeamSchematicHarmonicTot}: spatial-only modulation is shown in Fig.~\ref{Fig:H_S_UC} and corresponds to the case with $T_m\rightarrow\infty$, thus $\omega_m\rightarrow0$ and $\nu_m\rightarrow0$, while spatiotemporal modulation with $\alpha_m=0.20$ and $\nu_m\neq0$ is shown in Fig.~\ref{Fig:H_ST_UC}. 
The Fourier coefficients in Eq.~\eqref{eq:permatseries} are given by Eq.~\eqref{Eq:FourierCoeff} and write:
\begin{equation}
\hat{E}_p = \begin{cases} E_0 &\mbox{if } p = 0 \\ 
E_m/2 & \mbox{if } p = \pm1 \\  0 &\mbox{otherwise } \end{cases}
\end{equation}
while $\hat{\rho}_p=\rho_0\delta_{p0}$, where $\delta_{p0}$ is the Kronecker delta.

We also consider a square traveling wave of the Young's modulus only given by the following expressions:
\begin{equation}\label{Eq:SM}
E(x,t) = E_1 + (E_2 - E_1) H\big[\cos(\omega_m t - \kappa_m x)\big], \quad\quad\quad \rho(x,t) = \rho_0,
\end{equation}
in which $H$ denotes the Heaviside function. The Young's modulus takes the two values $E_1$ and $E_2$, with $E_2>E_1$. In this case, $E_m=E_2-E_1$. The spatial and time periods of the modulation are $\lambda_m=2\pi/\kappa_m$ and $T_m=2\pi/\omega_m$, respectively. The modulation travels with velocity $v_m=\omega_m/\kappa_m$.
It can be shown that the Fourier coefficients for square wave modulation given by Eq.~\eqref{Eq:FourierCoeff} write:
\begin{equation}\label{eq:FourierCoeff_SM}
\hat{E}_p = E_2 \delta_p + \frac{E_1 - E_2}{2} \sinc\Big[\frac{\pi p}{2}\Big]
\end{equation}
in which $\sinc(x)=sin(x)/x$ and $\hat{\rho}_p=\rho_0\delta_{p0}$. 
\subsection{Band diagrams}
\subsubsection{Harmonic modulation}
\begin{figure}[hbtp]
	\centering       
	\begin{subfigure}[b]{0.48\textwidth}
		\includegraphics[width=\textwidth]{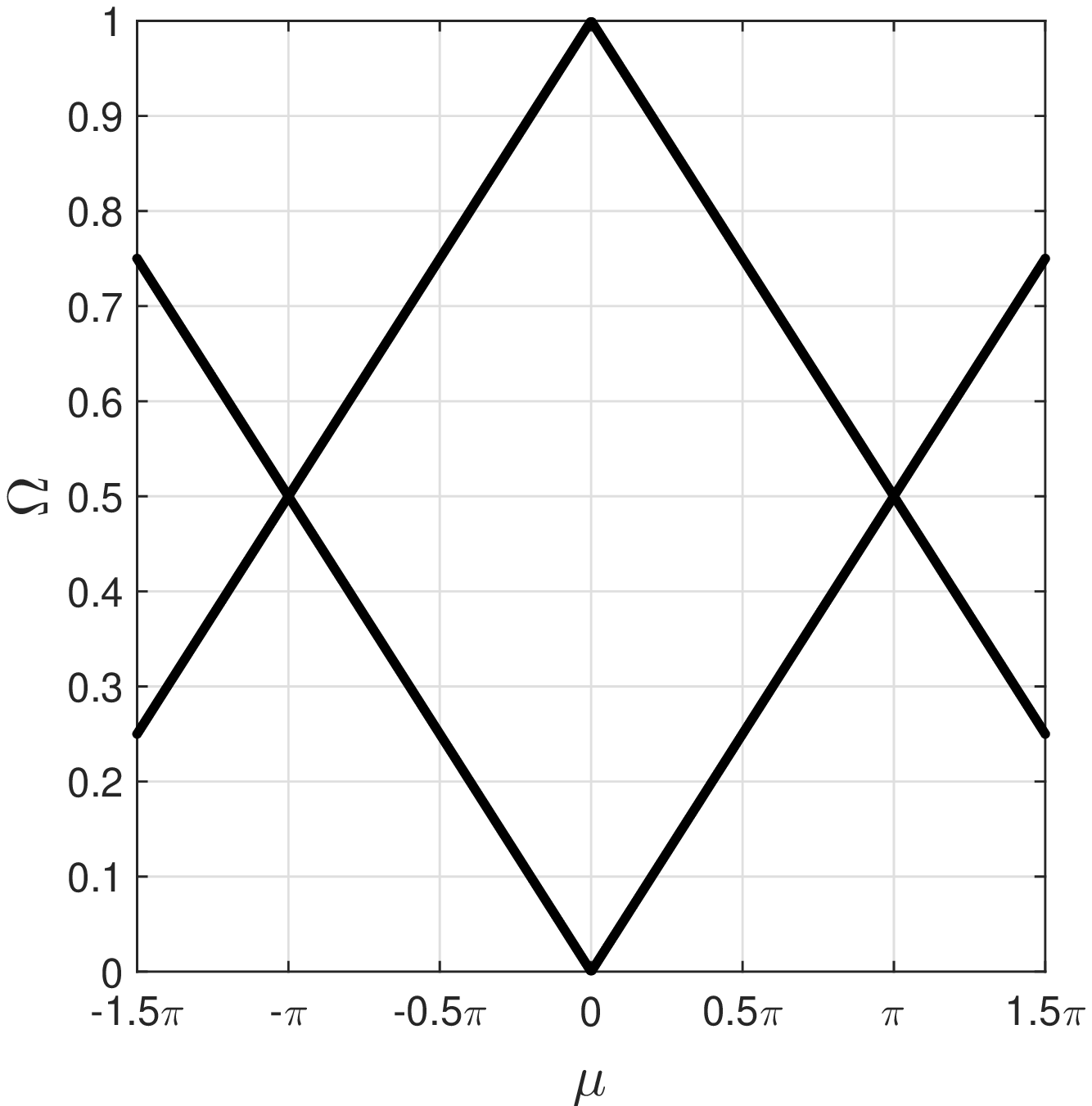}
		\caption{}
		\label{Fig:BD_R_H_01_BG_noNegFreq1dot5pi}
	\end{subfigure}               
	\begin{subfigure}[b]{0.48\textwidth}
		\includegraphics[width=\textwidth]{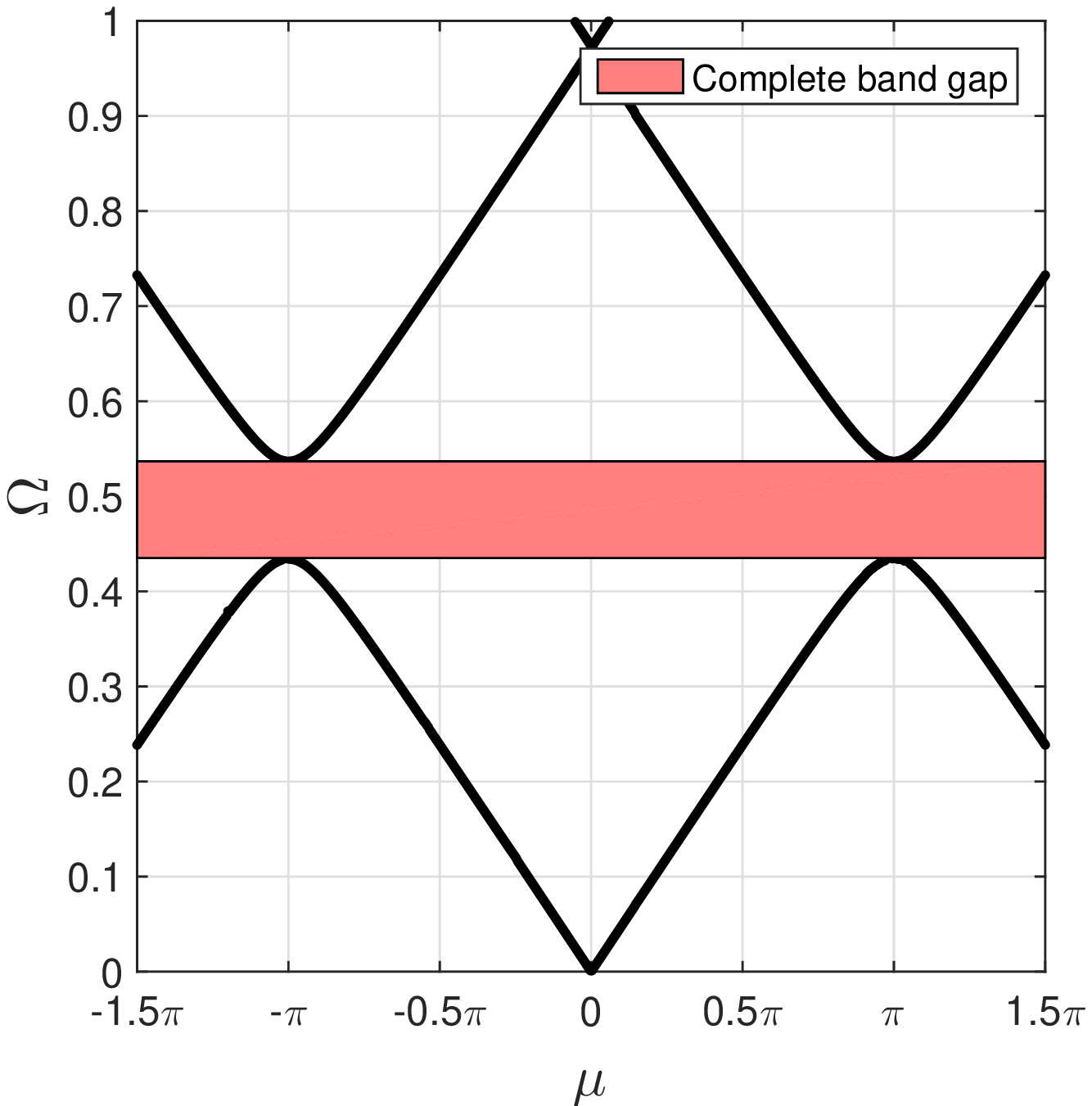}
		\caption{}
		\label{Fig:BD_R_H_02_BG_noNegFreq1dot5pi}
	\end{subfigure}
	
	\begin{subfigure}[b]{0.48\textwidth}
		\includegraphics[width=\textwidth]{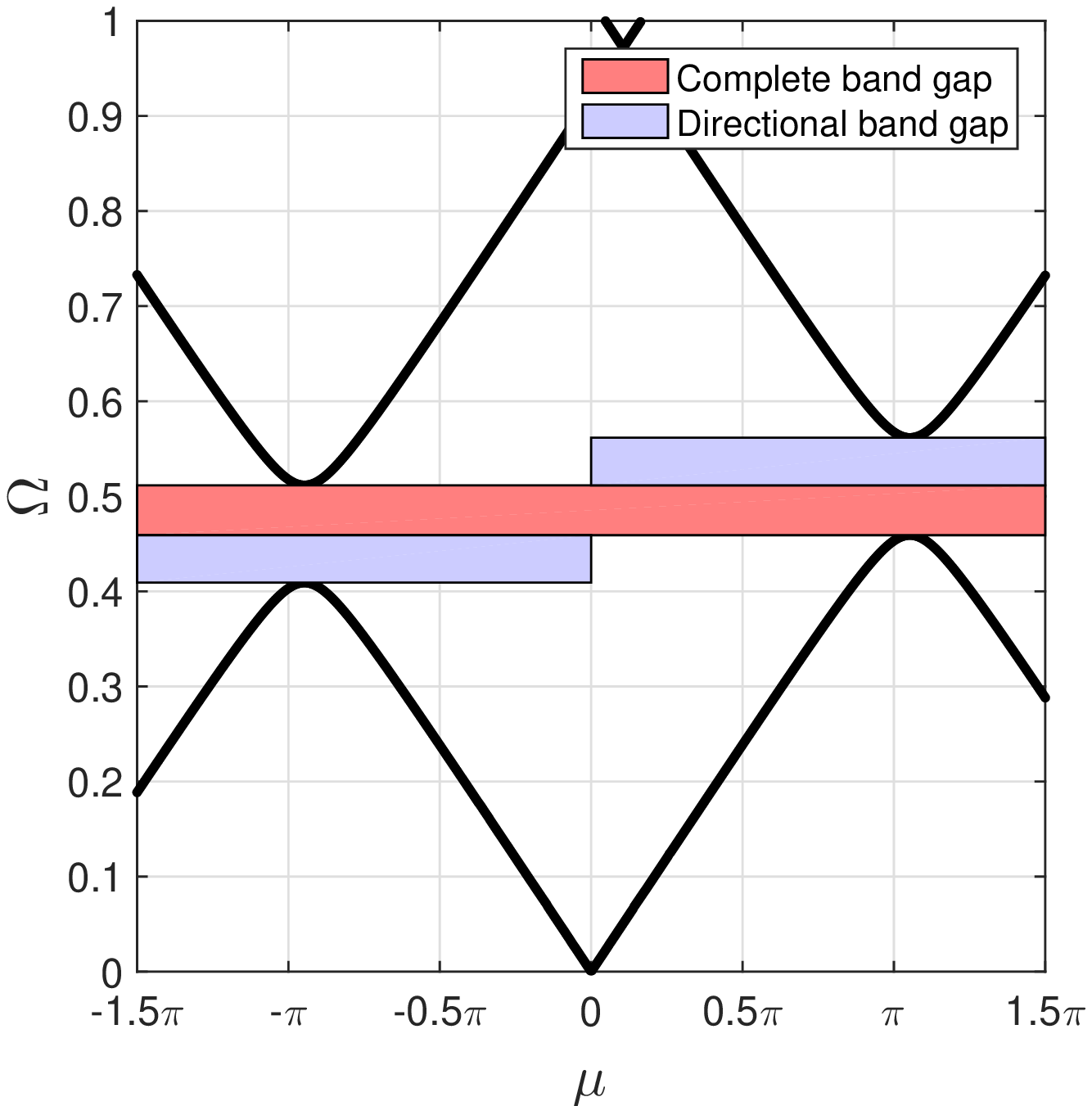}
		\caption{}
		\label{Fig:BD_R_H_03_BG_noNegFreq1dot5pi}
	\end{subfigure}               
	\begin{subfigure}[b]{0.48\textwidth}
		\includegraphics[width=\textwidth]{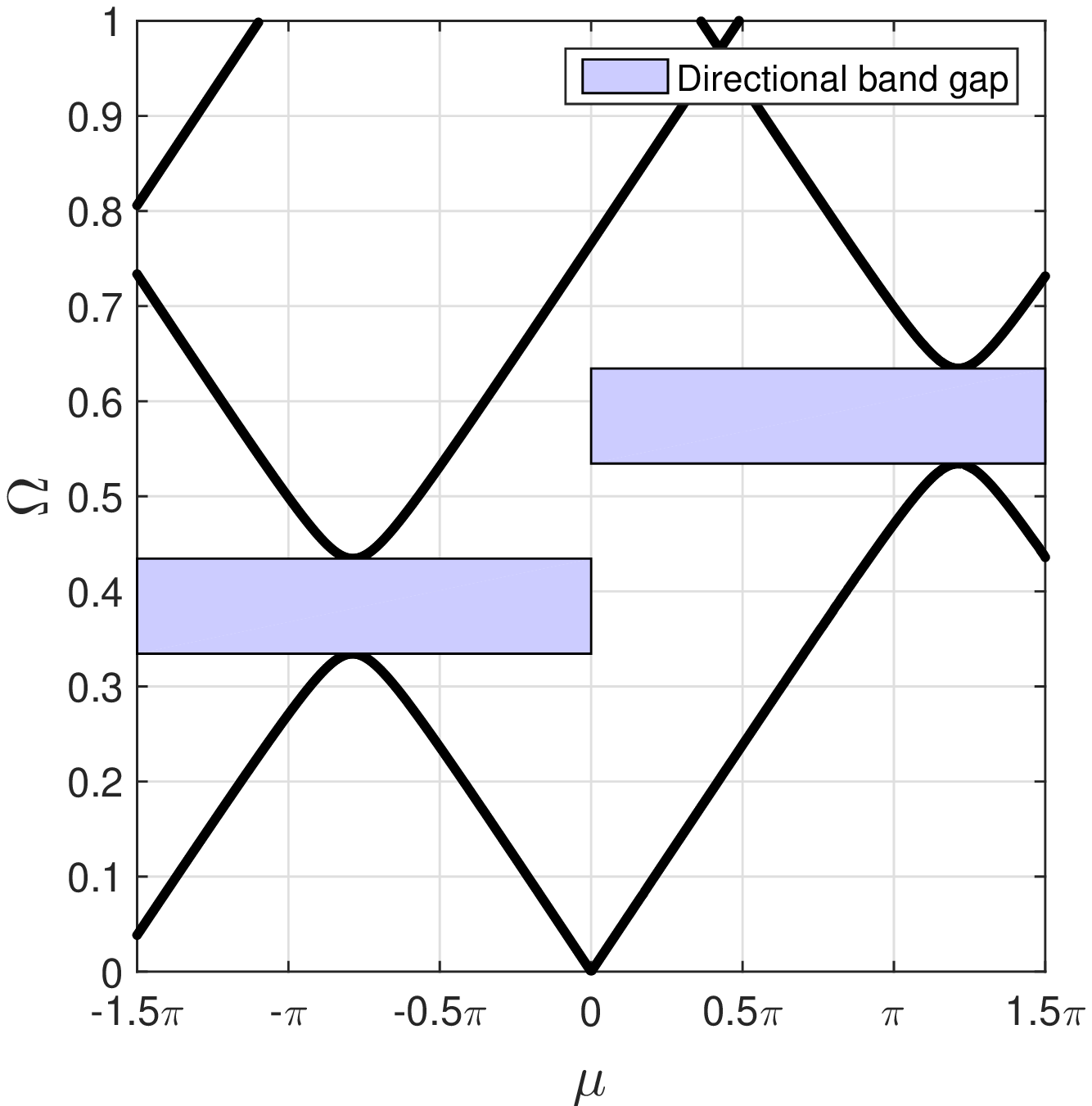}
		\caption{}
		\label{Fig:BD_R_H_04_BG_noNegFreq1dot5pi}
	\end{subfigure}
	\caption{Band diagrams for beam in longitudinal motion and harmonic modulation: (a) non-modulated beam with $\alpha_m =0$ and $\nu_m=0$, (b) space modulated only beam with $\alpha_m =0.40$ and $\nu_m=0$, (c) space-time modulated beam with $\alpha_m =0.40$ and $\nu_m=0.05$, (d) space-time modulated beam with $\alpha_m =0.40$ and $\nu_m=0.20$. For $\alpha_m \neq0$ and $\nu_m\neq0$ the mirror symmetry with respect to the frequency axis is relaxed, leading to directional band gaps which signal the possibility of one-way propagation.}
	\label{Fig:BD_R_H}
\end{figure}

\begin{figure}[hbtp]
	\centering       
	\begin{subfigure}[b]{0.48\textwidth}
		\includegraphics[width=\textwidth]{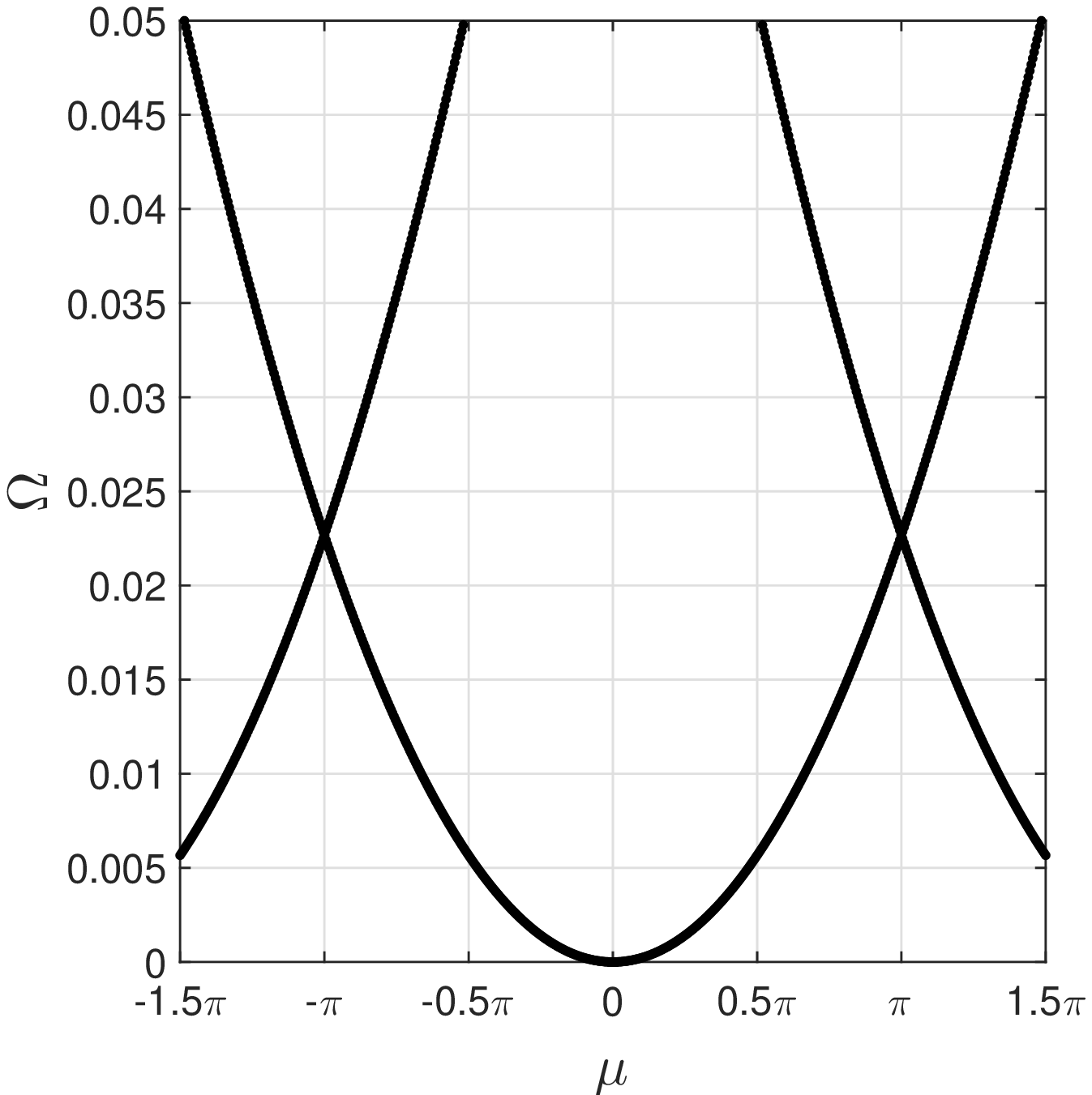}
		\caption{}
		\label{Fig:BD_B_H_01_BG_noNegFreq1dot5pi_Leg_Om0dot05}
	\end{subfigure}               
	\begin{subfigure}[b]{0.48\textwidth}
		\includegraphics[width=\textwidth]{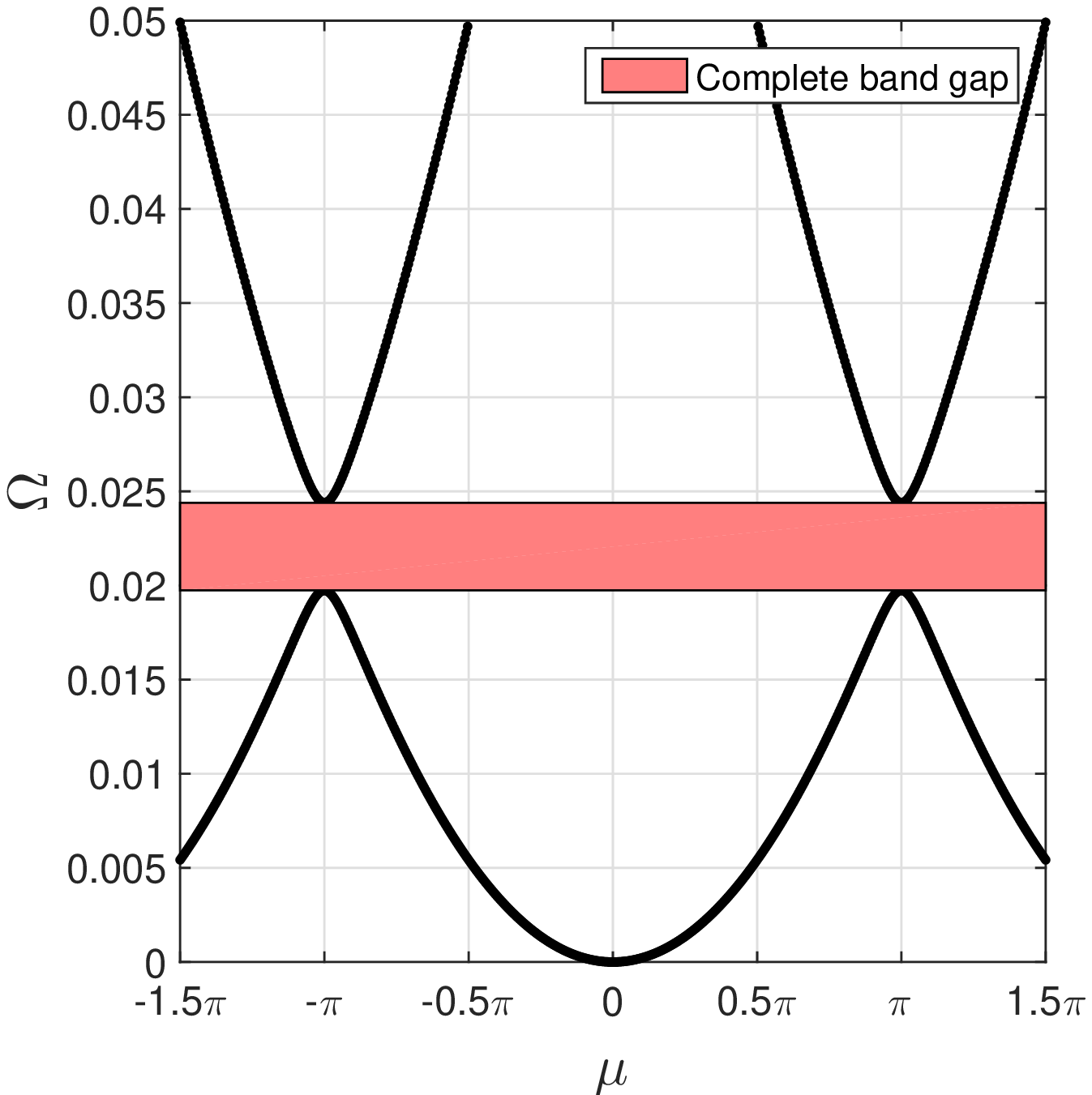}
		\caption{}
		\label{Fig:BD_B_H_02_BG_noNegFreq1dot5pi_Leg_Om0dot05}
	\end{subfigure}
	
	\begin{subfigure}[b]{0.48\textwidth}
		\includegraphics[width=\textwidth]{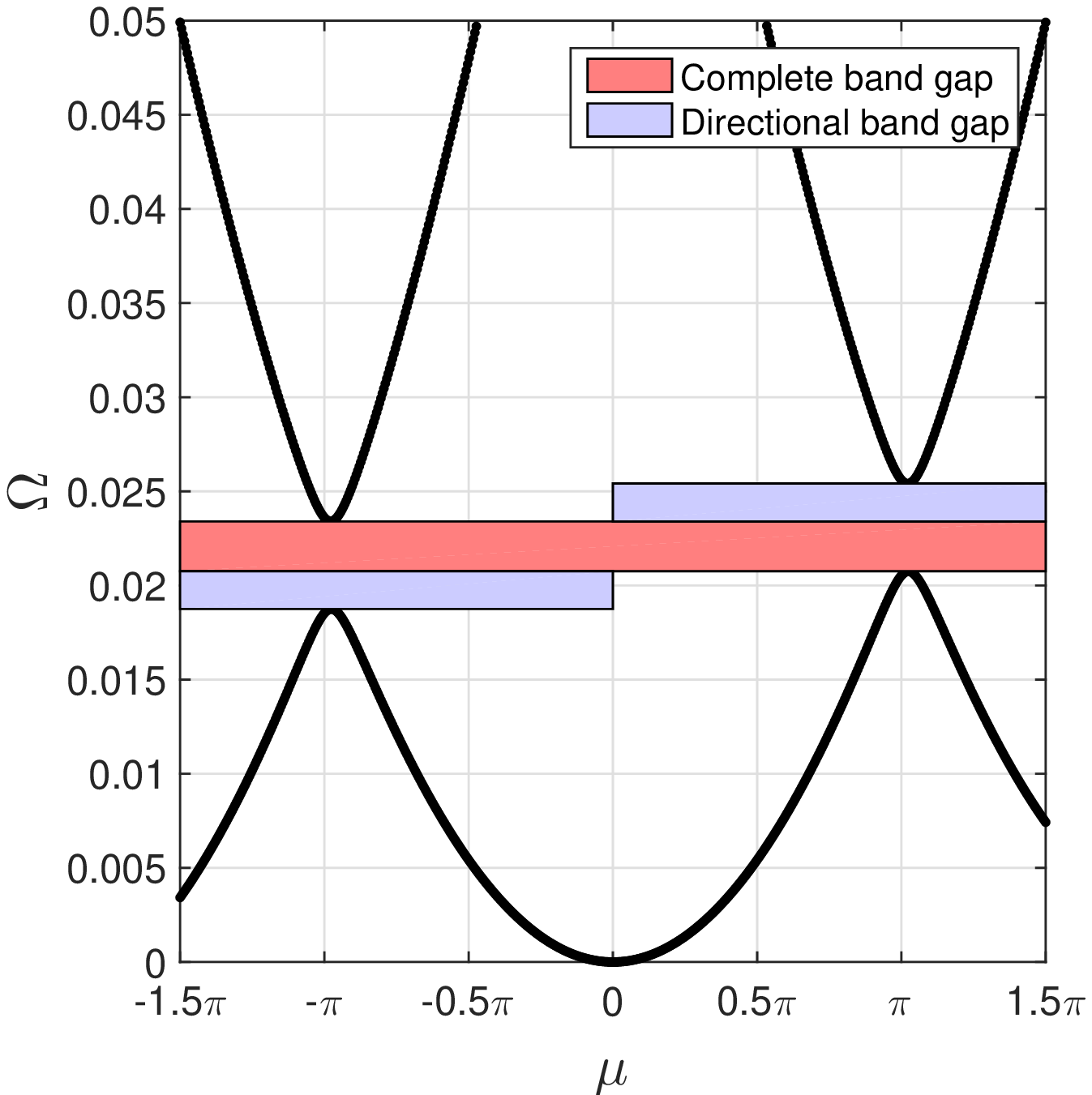}
		\caption{}
		\label{Fig:BD_B_H_03_BG_noNegFreq1dot5pi_Leg_Om0dot05}
	\end{subfigure}               
	\begin{subfigure}[b]{0.48\textwidth}
		\includegraphics[width=\textwidth]{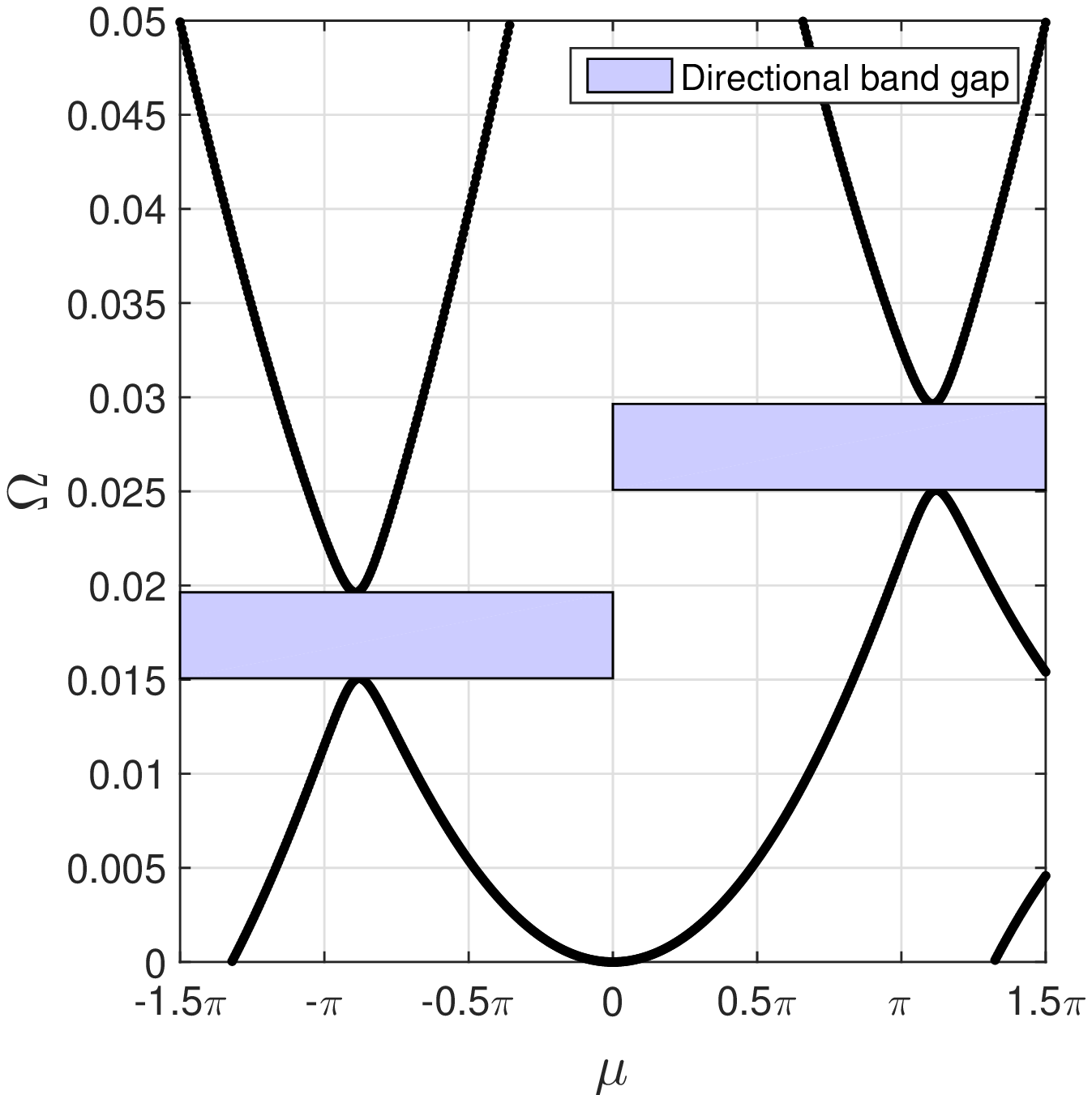}
		\caption{}
		\label{Fig:BD_B_H_04_BG_noNegFreq1dot5pi_Leg_Om0dot05}
	\end{subfigure}
	\caption{Band diagrams for beam in transverse motion and harmonic modulation: (a) non-modulated beam with $\alpha_m =0$ and $\nu_m=0$, (b) space modulated only beam with $\alpha_m =0.40$ and $\nu_m=0$, (c) space-time modulated beam with $\alpha_m =0.40$ and $\nu_m=0.002$, (d) space-time modulated beam with $\alpha_m =0.40$ and $\nu_m=0.01$. For $\alpha_m \neq0$ and $\nu_m\neq0$ the mirror symmetry is lost and directional band gaps are obtained.}
	\label{Fig:BD_B_H}
\end{figure}

We obtain the band diagrams shown in Fig.~\ref{Fig:BD_R_H} and Fig.~\ref{Fig:BD_B_H} by solving the QEP in Eq.~\eqref{eq:QEP_L} and Eq.~\eqref{eq:QEP_T} for longitudinal and transverse motion, respectively.
For conventional periodic system with spatial-only modulation, the dispersion diagrams are periodic in the wavenumber domain, therefore the calculation of the dispersion diagrams can be restricted to one period, known as First Brillouin Zone (FBZ), where the dimensionless wavenumber $\mu$ ranges in $[-\pi,\pi]$. Moreover, it is customary to further restrict the domain of the computation by considering just half of the period, called Irreducible Brillouin Zone (IBZ), with $\mu$ ranging in $[0,\pi]$. This approach is justified by an important property of conventional periodic structures, namely that their dispersion relation satisfies the relation $\Omega(\mu)=\Omega(-\mu)$. Such property is due to the fact that the structure can support both backward propagating waves and forward propagating waves, and the behavior of the waves is the same no matter which direction they are propagating to. It is also understood that the relation $\Omega(\mu)=\Omega(-\mu)$ implies symmetry of the dispersion diagrams with respect to the frequency axis, allowing the analysis to be restricted on the IBZ~\cite{brillouin1953wave}. We show that for spatiotemporal modulation of the material properties $\Omega(\mu)\neq\Omega(-\mu)$, therefore the band diagrams are not symmetric with respect to the frequency axis. For this reason, we need to consider a sufficiently broad range for the values of the wavenumber. In our analysis, we impose $\mu$ to range in $[-3/2\pi,3/2\pi]$. In all the computations, we focus on the two lowest dispersion branches where a band gap opens for $\alpha_m\neq0$, so we can truncate the series expressing the solution imposing $N=3$. Therefore, the matrices $\hat{\bL}_j$ and $\hat{\bT}_j$, with $j=0,1,2$, in the QEP are square matrices of order $2\cdot3+1=7$. Both for longitudinal and transverse motion in absence of modulation, thus for $\alpha_m =0$ and $\nu_m=0$, the beam does not display any band gap (Fig.~\ref{Fig:BD_R_H_01_BG_noNegFreq1dot5pi} and \ref{Fig:BD_B_H_01_BG_noNegFreq1dot5pi_Leg_Om0dot05}). The only dispersion branches having physical meaning are the ones departing from the origin, which are associated to the two $0-$th order harmonics, one associated to the backward-propagating wave, the other associated to the forward-propagating wave. On the contrary, when space-only modulation is applied, in our example for $\alpha_m =0.40$ and $\nu_m=0$, the system displays a complete band gap for $\Omega$ in the range $[0.43,0.53]$ for longitudinal motion and $[0.0197,0.0244]$ for transverse motion, in which both backward-propagating waves and forward-propagating waves are not supported by the structure, as shown in Fig.~\ref{Fig:BD_R_H_02_BG_noNegFreq1dot5pi} and Fig.~\ref{Fig:BD_B_H_02_BG_noNegFreq1dot5pi_Leg_Om0dot05}. Due to spatial periodic modulation of the medium, higher spatial-only harmonics in the solutions expressed by Eq.~\eqref{eq:sol1} and Eq.~\eqref{eq:sol2} have nonzero amplitude and coupling between the associated modes leads to  dispersive behavior and band gaps of Bragg type at the edge of the FBZ. The band diagram is periodic with period $2\pi$, with FBZ ranging from $\mu=-\pi$ to $\mu=\pi$. When modulation in space \textit{and} time is introduced, symmetry of the dispersion diagram with respect to $\Omega$ and $\mu$ is broken, as shown in Fig.~\ref{Fig:BD_R_H_03_BG_noNegFreq1dot5pi} and Fig.~\ref{Fig:BD_B_H_03_BG_noNegFreq1dot5pi_Leg_Om0dot05}, which are obtained respectively for $\alpha_m =0.40$ and $\nu_m=0.05$ for longitudinal motion, and for $\alpha_m =0.40$ and $\nu_m=0.002$ for transverse motion. The band diagrams for spatiotemporal periodic structures clearly show that, within certain frequency ranges, only modes propagating in one direction are allowed. For longitudinal motion, a closer look at Fig.~\ref{Fig:BD_R_H_03_BG_noNegFreq1dot5pi} reveals that waves propagating with $\Omega$ in the range $[0.41,0.46]$ have positive group velocity only, therefore forward propagating waves only are allowed to propagate, while backward-propagating longitudinal waves only are supported by the structure for $\Omega$ in the range $[0.51,0.56]$. Such frequency ranges are then associated to one-way propagation, thus we can call them ``directional band gaps''. By comparing the band diagrams for longitudinal motion in Fig.~\ref{Fig:BD_R_H_03_BG_noNegFreq1dot5pi} and~\ref{Fig:BD_R_H_04_BG_noNegFreq1dot5pi}, we also observe that for $\alpha_m=0.40$ and $\nu_m=0.05$ the structure displays a complete band gap for $\Omega$ in the range $[0.46,0.51]$ together with the two directional band gaps for $\Omega$ in the ranges $[0.41,0.46]$ and $[0.51,0.56]$, while for $\alpha_m=0.40$ and $\nu_m=0.20$ the structure exhibit just two wider directional band gaps. 
Analogous considerations can be done for transverse motion by analyzing the dispersion diagrams in Fig.~\ref{Fig:BD_B_H}. Furthermore, Fig.~\ref{Fig:BD_R_H_03_BG_noNegFreq1dot5pi} and Fig.~\ref{Fig:BD_R_H_04_BG_noNegFreq1dot5pi} reveal that when $\nu_m\neq0$, the local maxima of the dispersion branches do not coincide with the limits of the classically defined FBZ $\mu=\pm\pi$, instead they are shifted in the direction of positive wavenumbers $\kappa$ for forward propagating modulation. Similar considerations hold in the case of transverse motion, as shown in Fig.~\ref{Fig:BD_B_H_03_BG_noNegFreq1dot5pi_Leg_Om0dot05} and ~\ref{Fig:BD_B_H_04_BG_noNegFreq1dot5pi_Leg_Om0dot05}. In other words, spatiotemporal modulation of the material properties introduces shifts in the position of the band gaps, breaking symmetry of the dispersion diagrams and leading to directional band gaps. In addition, spatiotemporal modulation also challenges the classic definition of the FBZ and its limits, as already observed by Cassedy~\cite{Cassedy1963,Cassedy1965}.

\begin{figure}[hbtp]
	\centering       
	\begin{subfigure}[b]{0.48\textwidth}
		\includegraphics[width=\textwidth]{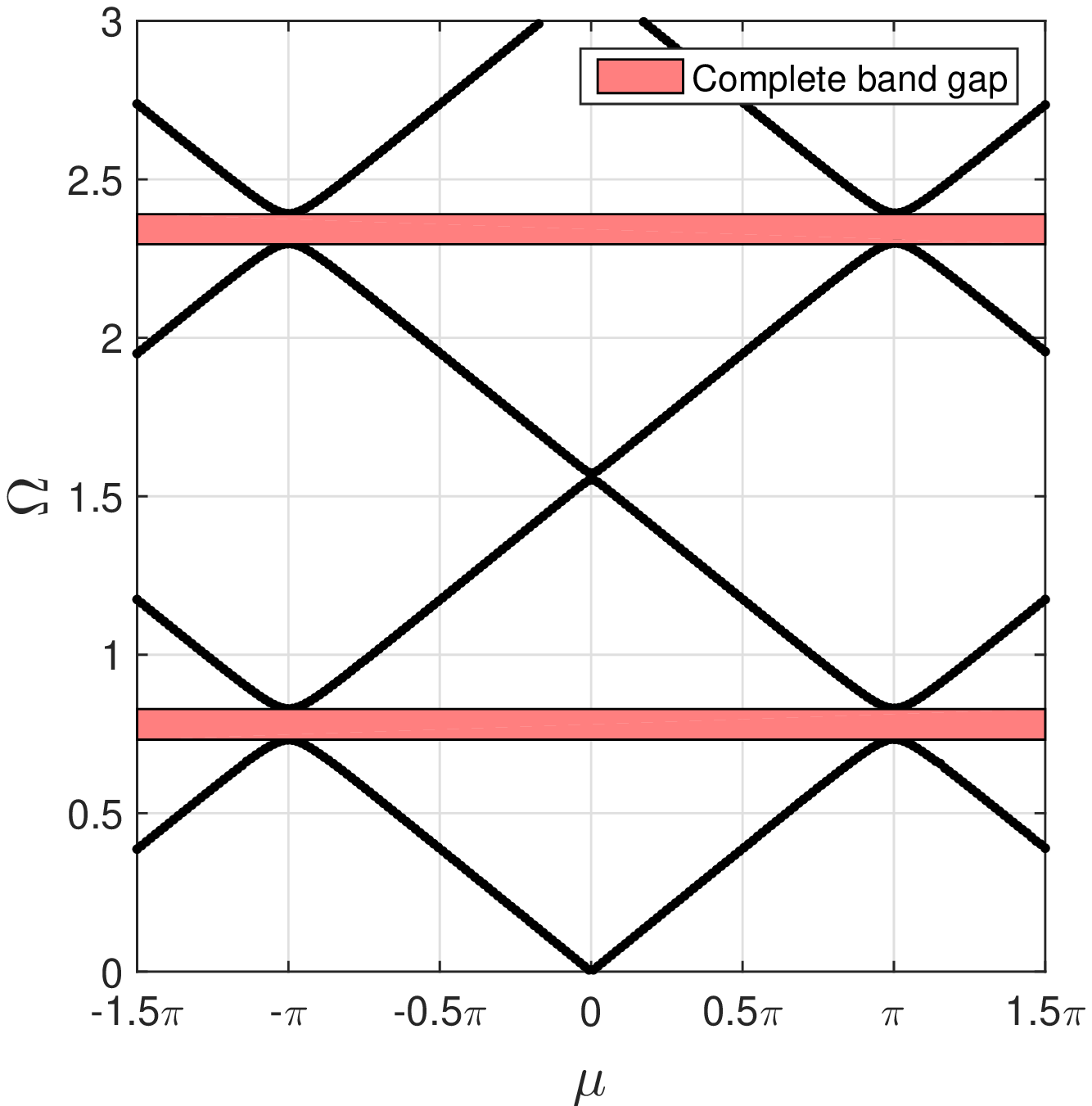}
		\caption{}
		\label{Fig:BD_R_S_01_BG_noNegFreq1dot5pi_Leg_Om3}
	\end{subfigure}               
	\begin{subfigure}[b]{0.48\textwidth}
		\includegraphics[width=\textwidth]{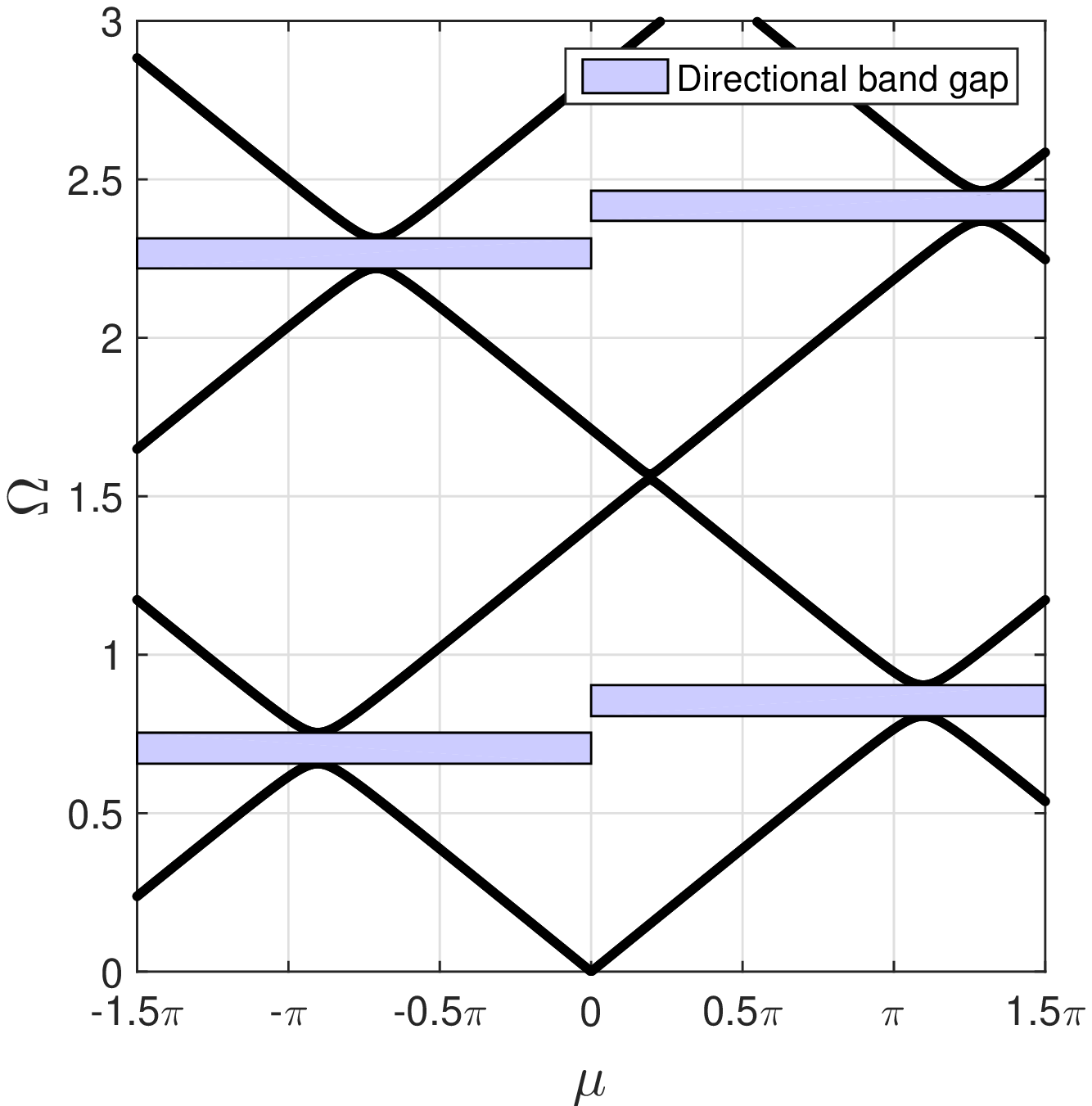}
		\caption{}
		\label{Fig:BD_R_S_02_BG_noNegFreq1dot5pi_Leg_Om3}
	\end{subfigure}      
	\begin{subfigure}[b]{0.48\textwidth}
		\includegraphics[width=\textwidth]{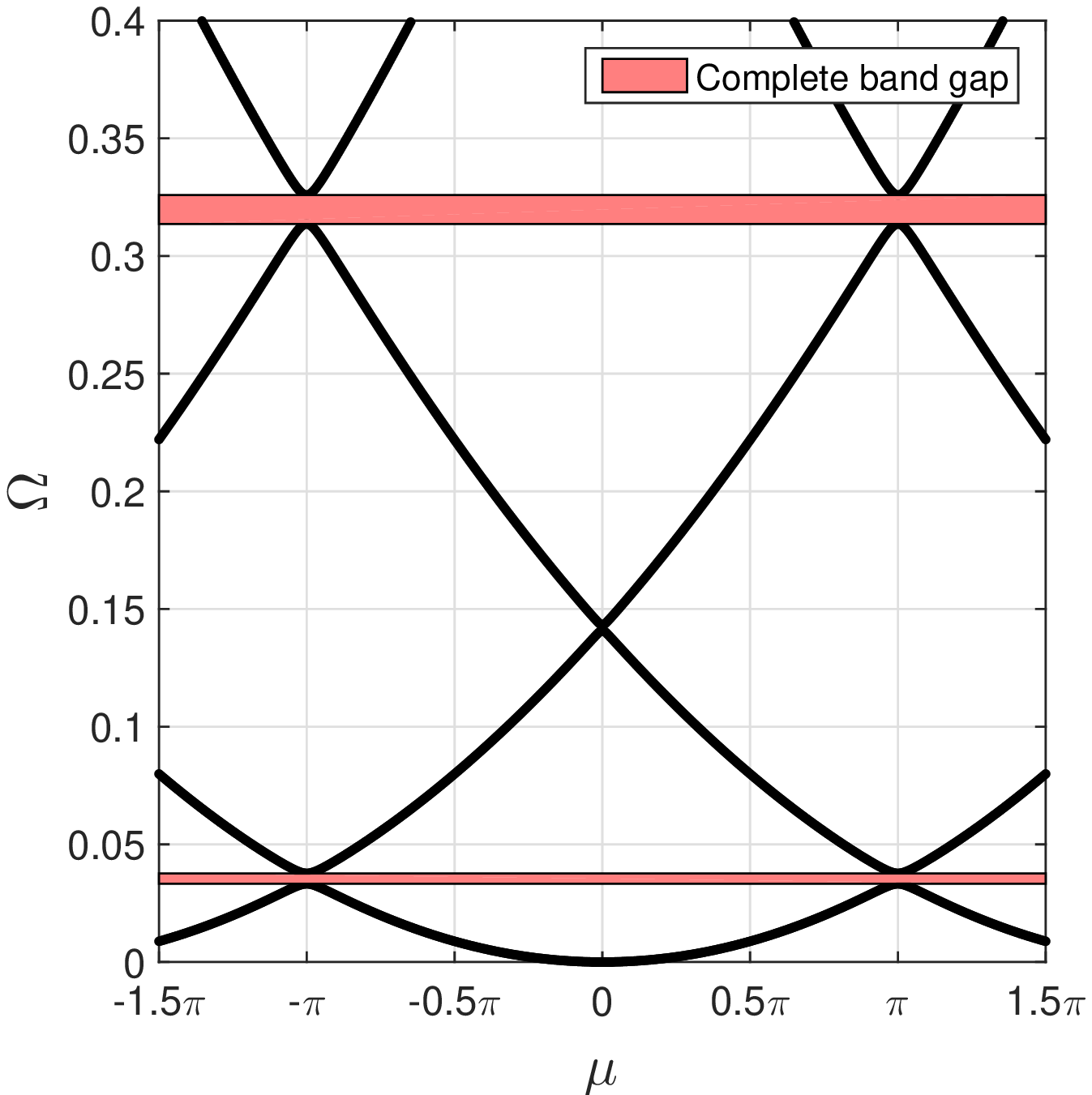}
		\caption{}
		\label{Fig:BD_B_S_01_BG_noNegFreq1dot5pi_Leg_Om0dot4}
	\end{subfigure}               
	\begin{subfigure}[b]{0.48\textwidth}
		\includegraphics[width=\textwidth]{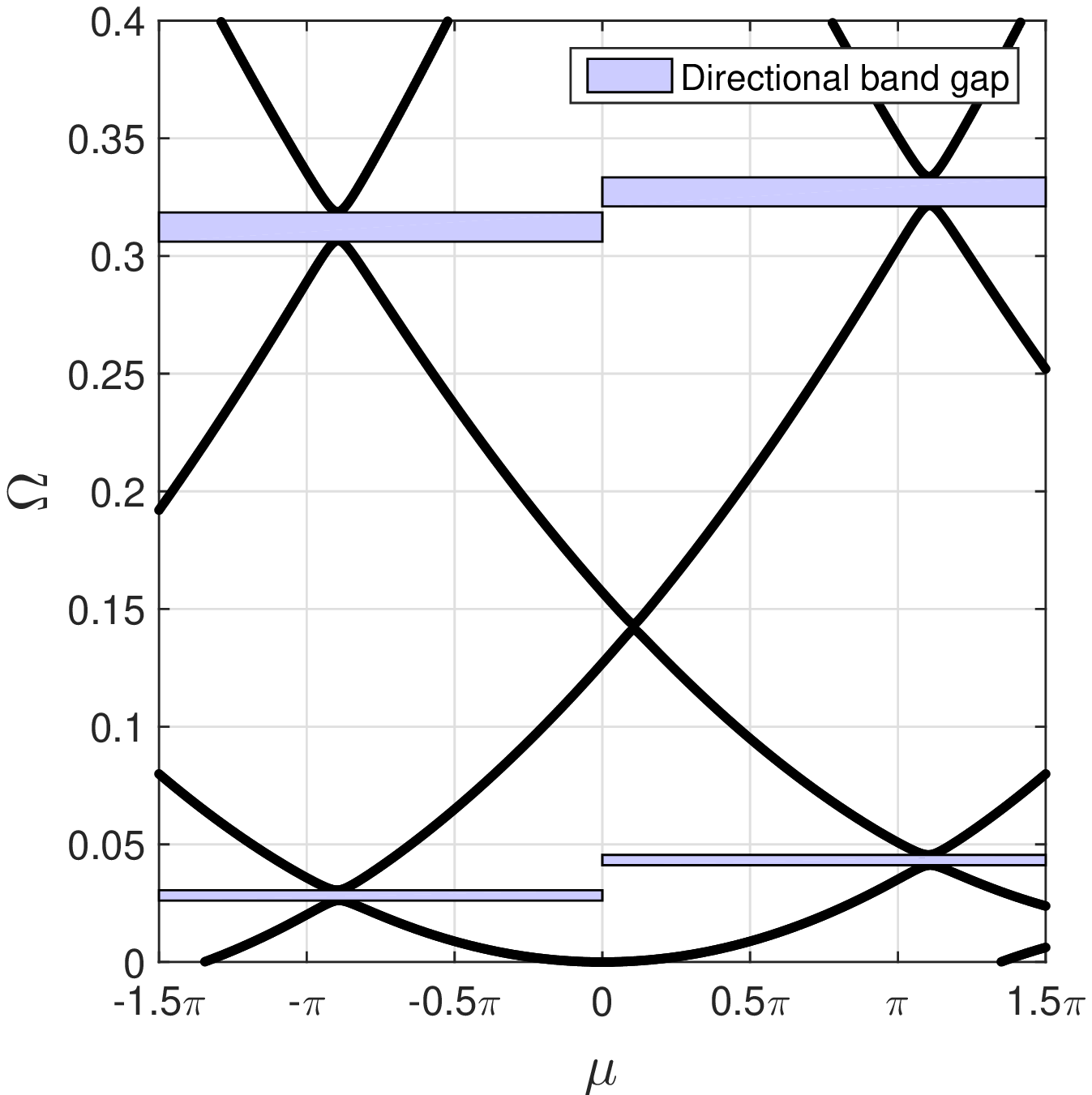}
		\caption{}
		\label{Fig:BD_B_S_02_BG_noNegFreq1dot5pi_Leg_Om0dot4}
	\end{subfigure}
	\caption{Band diagrams for beam in longitudinal (a,b) and transverse (c,d) motion and square modulation. For longitudinal motion: (a) space modulated only beam with $\alpha_m =3$ and $\nu_m=0$, (b) space-time modulated with $\alpha_m =3$ and $\nu_m=0.15$. Similarly for transverse motion: (c) space modulated only beam with $\alpha_m =3$ and $\nu_m=0$, (d) space-time modulated with $\alpha_m =3$ and $\nu_m=0.015$. The structure is non-reciprocal for $\alpha_m \neq0$ and $\nu_m\neq0$ in multiple frequency ranges.}
	\label{Fig:BD_B_S}
\end{figure}
\subsubsection{Square modulation}
For square modulation we also compute the band diagrams for both longitudinal and transverse motion and compare the cases of spatial-only modulation and spatiotemporal modulation. We consider $\alpha=3$ and $\nu_m=0.15$ for the beam in longitudinal motion, while for transverse motion we consider the case with $\alpha=3$ and $\nu_m=0.015$. In solving the QEP, we consider $N=5$. As shown in Fig.~\ref{Fig:BD_R_S_01_BG_noNegFreq1dot5pi_Leg_Om3} and Fig.~\ref{Fig:BD_B_S_01_BG_noNegFreq1dot5pi_Leg_Om0dot4}, the spatial-only modulation, thus for $\nu_m=0$, induces multiple band gaps in the structure. From Fig.~\ref{Fig:BD_R_S_02_BG_noNegFreq1dot5pi_Leg_Om3} and Fig.~\ref{Fig:BD_B_S_02_BG_noNegFreq1dot5pi_Leg_Om0dot4}, we can see that when spatiotemporal modulation is considered, in this case for $\nu_m\neq0$, the band gaps are shifted upwards or downwards depending on the direction of propagation, hence they become directional band gaps and inform on the non-reciprocal behavior of the structure.
\subsection{Parametric analysis for harmonic modulation}
\begin{figure}[hbtp]
	\centering
	\begin{subfigure}[b]{0.48\textwidth}
		\includegraphics[width=\textwidth]{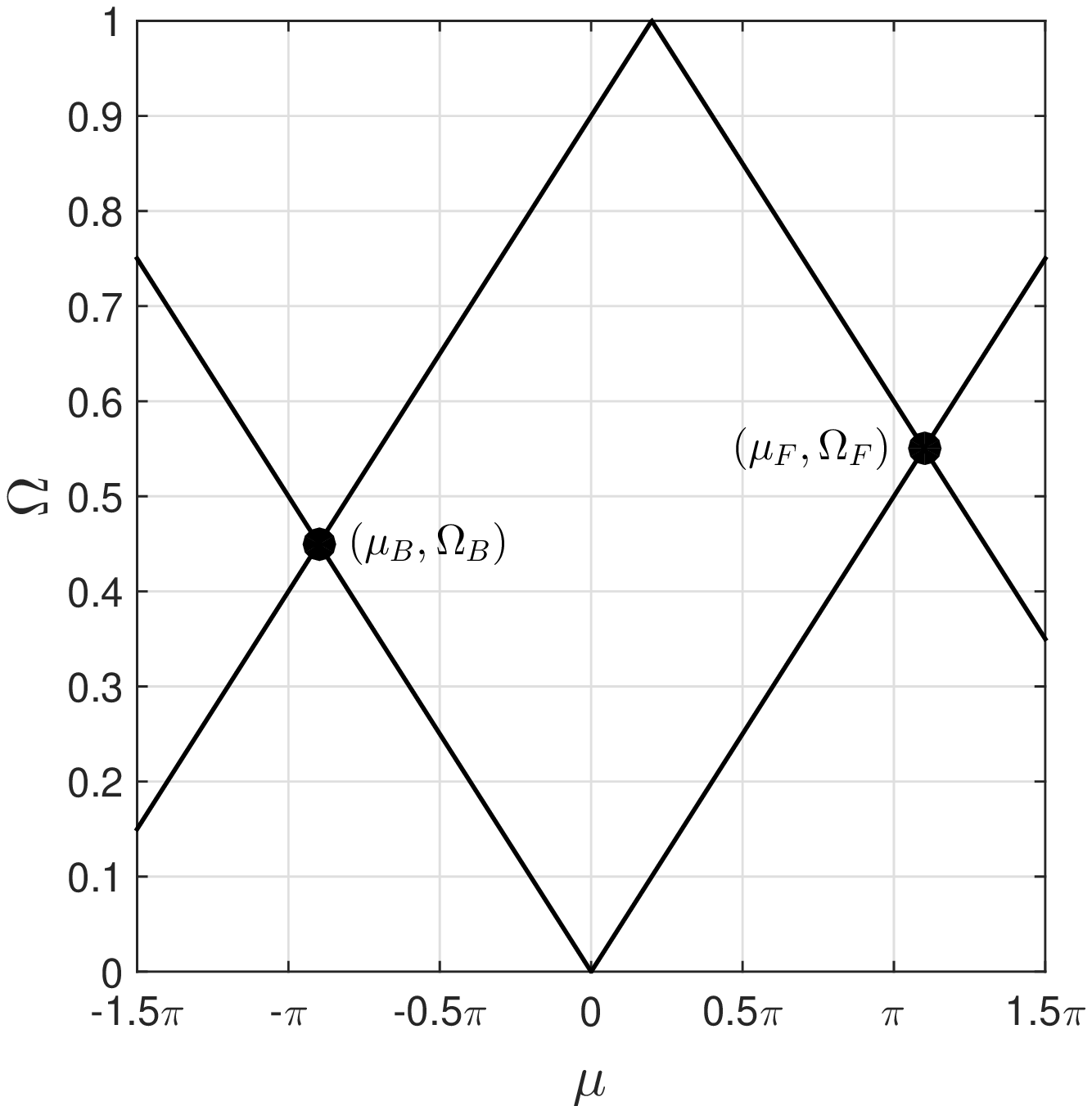}
		\caption{}
		\label{Fig:mu_F_mu_B_rod_tot}
	\end{subfigure}
	
	\begin{subfigure}[b]{0.48\textwidth}
		\includegraphics[width=\textwidth]{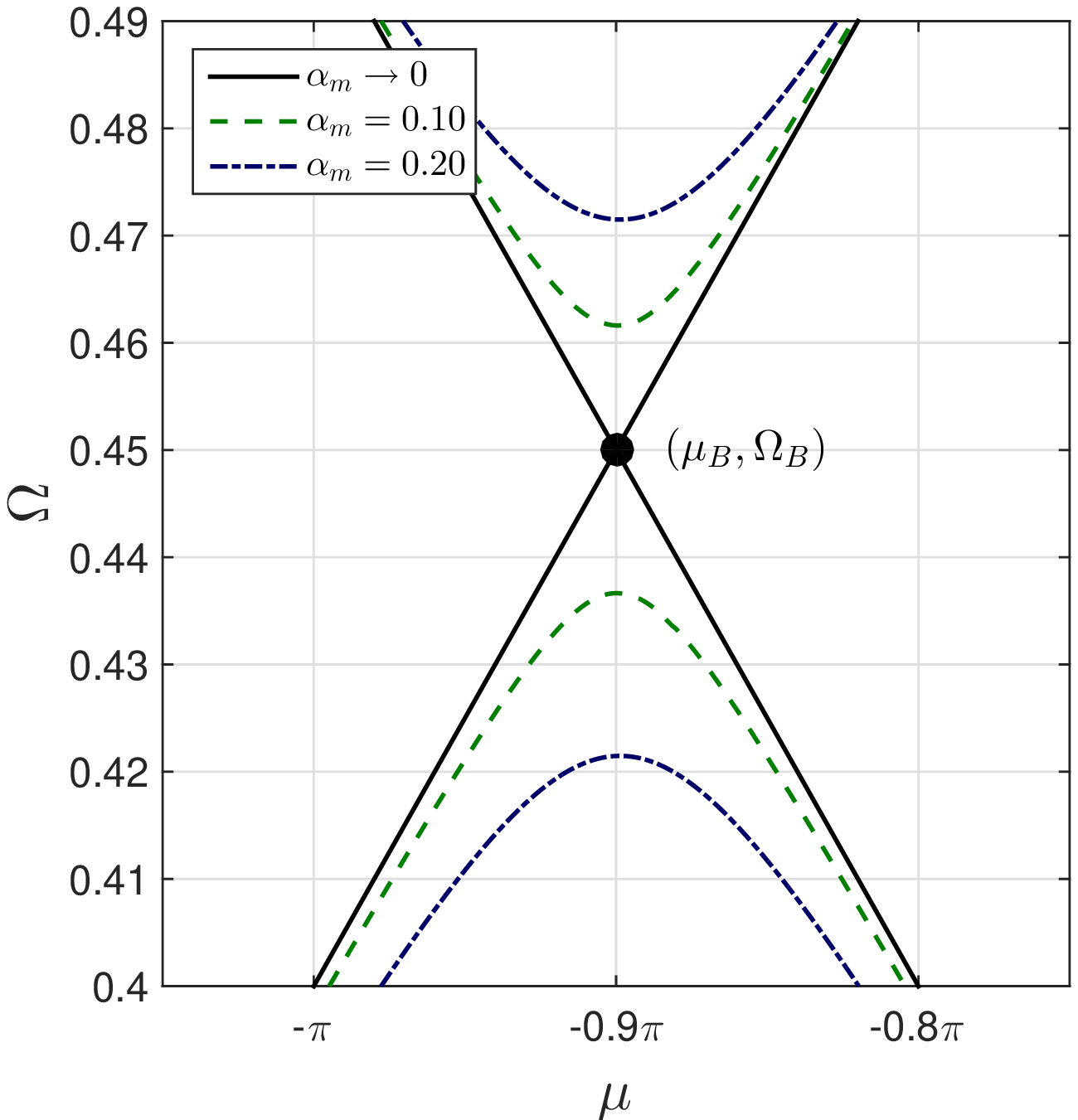}
		\caption{}
		\label{Fig:mu_B_rod}
	\end{subfigure}	
		\centering
	\begin{subfigure}[b]{0.48\textwidth}
		\includegraphics[width=\textwidth]{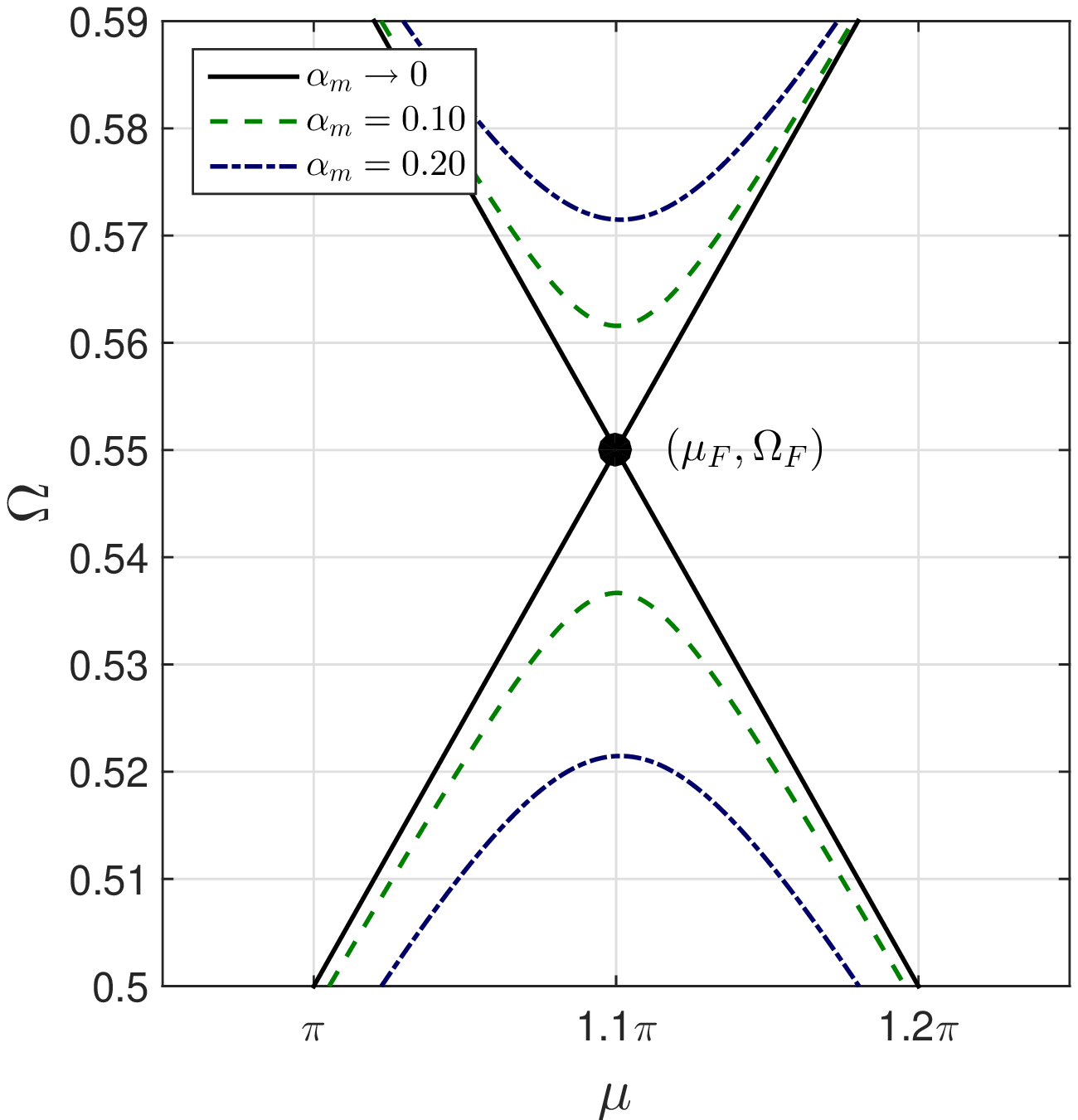}
		\caption{}
		\label{Fig:mu_F_rod}
	\end{subfigure}                   
	\caption{Band diagrams for longitudinal motion with $\nu_m=0.10$ and different values of $\alpha_m$. For $\alpha_m\to0$ (a), the first two dispersion branches intersect at $(\mu_B,\Omega_B)$ and $(\mu_F,\Omega_F)$ for backward and forward propagating waves, respectively. Comparison with the case in which $\alpha_m\neq0$ is given in (b) for backward propagating  waves and in (c) forward propagating waves.}
	\label{Fig:Analytical_rod}
\end{figure}

We now analytically characterize the effect of the the modulation parameters on the dispersion properties of the beam. We consider a reduced version of the QEP used to compute the band diagrams in Fig.~\ref{Fig:BD_R_H} and Fig.~\ref{Fig:BD_B_H} and obtain approximate analytical expressions for the dispersion branches of spatiotemporal modulated beams. Such expressions allow us to quantify the effect of the modulation parameters $\alpha_m$ and $\nu_m$ not only on the position and width of the directional band gaps, but also on the shift of the limits of the FBZ. We confine our analysis to the harmonic modulation only. We start with the longitudinal motion case by investigating the values of the FBZ limits induced by the modulation. Fig.~\ref{Fig:Analytical_rod} shows a detail of the regions of the band diagram associated with the directional band gaps. The plot presents different cases of modulation, each corresponding to the same value $\nu_m=0.10$ and increasingly smaller values of $\alpha_m$. In Fig.~\ref{Fig:mu_B_rod} and~\ref{Fig:mu_F_rod} we observe that for $\alpha_m\to0$, the band gaps close and two dispersion branches intersect at the point $(\mu_F,\Omega_F)$ for forward propagating waves, with $\mu_F\neq\pi$, and at the point $(\mu_B,\Omega_B)$ for backward propagating waves, with $\mu_B\neq-\pi$.
We seek for analytic expressions for $\mu_B$, $\Omega_B$ and $\mu_F$, $\Omega_F$ by truncating the series expressing the solution in Eq.~\eqref{eq:sol1} to the first order, then Eq.~\eqref{eq:QEP_L} reduces to:
\begin{equation}\label{eq:redQEP}
\begin{bmatrix} 
L_{11} & L_{12}  & L_{13}  \\ 
L_{21} & L_{22}  & L_{23}  \\ 
L_{31} & L_{32}  & L_{33} 
\end{bmatrix} 
\begin{bmatrix} 
\hat{u}_{-1}   \\ 
\hat{u}_{0}    \\ 
\hat{u}_{+1}   
\end{bmatrix} =
\begin{bmatrix} 
0  \\ 
0  \\ 
0   
\end{bmatrix}
\end{equation}
where the coefficient are expressed as:
\begin{align}
L_{11} &= \Omega^2 - 2 \nu_m \Omega - \Big[\Big(\frac{\mu}{2 \pi}-1\Big)^2 - \nu_m^2\Big] \\
L_{12} &=  -\frac{\mu}{2 \pi}\Big(\frac{\mu}{2 \pi}-1\Big)\frac{\alpha_m}{2} \\
L_{13} &=  0 \\
L_{21} &=    -\frac{\mu}{2 \pi}\Big(\frac{\mu}{2 \pi}-1\Big)\frac{\alpha_m}{2} \\
L_{22} &=  \Omega^2 - \Big(\frac{\mu}{2 \pi}\Big)^2 \\
L_{23} &=    -\frac{\mu}{2 \pi}\Big(\frac{\mu}{2 \pi}+1\Big)\frac{\alpha_m}{2} \\
L_{31} &=  0 \\
L_{32} &=   -\frac{\mu}{2 \pi}\Big(\frac{\mu}{2 \pi}+1\Big)\frac{\alpha_m}{2} \\
L_{33} &=  \Omega^2 + 2 \nu_m \Omega - \Big[\Big(\frac{\mu}{2 \pi}+1\Big)^2 - \nu_m^2\Big]
\end{align}
We first focus on the directional band gap for forward propagating waves. In this case, it can be shown that we can further simplify Eq.~\eqref{eq:redQEP} by solving the problem for the coefficients $\hat{u}_{-1}$ and $\hat{u}_{0}$ only to obtain:
\begin{equation}\label{eq:redQEP_forward}
\begin{bmatrix} 
L_{11} & L_{12}    \\ 
L_{21} & L_{22}  

\end{bmatrix} 
\begin{bmatrix} 
\hat{u}_{-1}   \\ 
\hat{u}_{0}    
\end{bmatrix} =
\begin{bmatrix} 
0  \\ 
0   
\end{bmatrix}
\end{equation}
A compact solution to this QEP is obtained by observing that for $\alpha_m\to0$, then $L_{12},L_{21}\to0$. The characteristic equation for Eq.~\eqref{eq:redQEP_forward} reduces to:
\begin{equation}
L_{11}L_{22}= \bigg\{  \Omega^2 - 2 \nu_m \Omega - \Big[\Big(\frac{\mu}{2 \pi}-1\Big)^2 - \nu_m^2\Big] \bigg\} \bigg\{\Omega^2 - \Big(\frac{\mu}{2 \pi}\Big)^2 \bigg\}=0
\end{equation}
whose solutions are:
\begin{align}
\Omega_{1,2}=&\nu_m \pm \Big(\frac{\mu}{2 \pi}-1\Big)\\
\Omega_{3,4}=&\pm\frac{\mu}{2 \pi}
\end{align}
Each of these four roots is associated to a different dispersion branch. In general these roots are distinct, but at $\mu=\mu_F$ Eq.~\eqref{eq:redQEP_forward} allows double roots for $\alpha_m\to0$. Two of the dispersion branches intersect at the point $(\mu_F,\Omega_F)$, which can be obtained by imposing:
\begin{equation}
\frac{\mu_F}{2 \pi}=\nu_m - \Big(\frac{\mu_F}{2 \pi}-1\Big)
\end{equation}
to get:
\begin{equation}\label{eq:muFOmF_rod}
\mu_F = \pi (1+\nu_m ) \qquad \qquad \Omega_F =\frac{1}{2}(1+\nu_m )
\end{equation}
These expressions show that the right limit of the FBZ is given by $\mu=\pi$ only if $\nu_m=0$, otherwise the limit of the FBZ between the first and the second dispersion branches is shifted by the value $\Delta \mu=\nu_m \pi$, which in first approximation depends upon the modulation speed only. 
If the dimensionless modulation amplitude is small but different from zero, $\alpha_m<<1$, the characteristic equation for Eq.~\eqref{eq:redQEP_forward} is a quartic and does not allow double roots at $\mu_F = \pi (1+\nu_m )$. Instead, it allows four distinct solutions, two of which lie in a neighborhood of $\Omega_F$. Therefore we can assume that:
\begin{equation}
L_{11} - L_{22} \approx 0
\end{equation}
for $\mu = \mu_F=\pi (1+\nu_m )$, in which equality holds for $\alpha_m\to0$. The latter key approximation allows us to study two simple problems:
\begin{align}\label{eq:Approx1_Rod}
L_{11}^2-L_{12}^2=&0 \qquad \rightarrow (L_{11}-L_{12})(L_{11}+L_{12})=0\\
L_{22}^2-L_{12}^2=&0 \qquad \rightarrow (L_{22}-L_{12})(L_{22}+L_{12})=0\label{eq:Approx2_Rod}
\end{align}
both representing an approximation of the original problem expressed by the characteristic equation of Eq.~\eqref{eq:redQEP_forward} in a neighborhood of ($\mu_F$,$\Omega_F$). The solutions of Eq.~\eqref{eq:Approx1_Rod} write:
\begin{equation}
\Omega_{1,2}=\alpha_m\pm\frac{1-\nu_m}{2}\sqrt{1\pm\frac{\alpha_m}{2}\frac{1+\nu_m}{1-\nu_m}}
\end{equation}
while the solutions of Eq.~\eqref{eq:Approx2_Rod} write:
\begin{equation}
\Omega_{3,4}=\pm\frac{1+\nu_m}{2}\sqrt{1\pm\frac{\alpha_m}{2}\frac{1-\nu_m}{1+\nu_m}}
\end{equation}
Each of the two approximate problems gives four roots, so by considering the positive values of frequency only, we define the following expressions:
\begin{align}\label{eq:OmegaTopBot_Rod_F}
\Omega^{top}_F =& \frac{1}{2}\bigg[\nu_m+\frac{1-\nu_m}{2}\sqrt{1+\frac{\alpha_m}{2}\frac{1+\nu_m}{1-\nu_m}}+\frac{1+\nu_m}{2}\sqrt{1+\frac{\alpha_m}{2}\frac{1-\nu_m}{1+\nu_m}}\bigg]\\
\Omega^{bot}_F =& \frac{1}{2}\bigg[\nu_m+\frac{1-\nu_m}{2}\sqrt{1-\frac{\alpha_m}{2}\frac{1+\nu_m}{1-\nu_m}}+\frac{1+\nu_m}{2}\sqrt{1-\frac{\alpha_m}{2}\frac{1-\nu_m}{1+\nu_m}}\bigg]
\end{align}
where $\Omega^{top}_F$ and $\Omega^{bot}_F$ are the upper and the lower limits of the directional band gap for the forward propagating waves, respectively.

Since $\alpha_m<<1$, linearized expressions can be obtained by using the Taylor series of $\sqrt{1 + x}=1+1/2x+O(x^2)$, thus Eq.~\eqref{eq:OmegaTopBot_Rod_F} rewrites as:
\begin{align}
\Omega^{top}_{F,lin}=&+\frac{\alpha_m}{8} +	\frac{1}{2}\Big(1+\nu_m\Big)\\
\Omega^{bot}_{F,lin}=&-\frac{\alpha_m}{8} +	\frac{1}{2}\Big(1+\nu_m\Big)
\end{align}
The dependency of the directional band gap width on the modulation parameters $\alpha_m$ and $\mu_m$ is given by the following expression:
\begin{equation}
\Delta\Omega_{F,lim}=\Omega^{top}_{F,lin}-\Omega^{bot}_{F,lin}=\frac{\alpha_m}{4}
\end{equation}
We can use a similar approach to study the directional band gap associated to the backward propagating waves. In this case, the terms $\hat{u}_0$ and $\hat{u}_{+1}$ are retained in Eq.~\eqref{eq:redQEP} to give:
\begin{equation}\label{eq:redQEP_backward}
\begin{bmatrix} 
L_{22} & L_{23}    \\ 
L_{32} & L_{33}  

\end{bmatrix} 
\begin{bmatrix} 
\hat{u}_{0}   \\ 
\hat{u}_{+1}    
\end{bmatrix} =
\begin{bmatrix} 
0  \\ 
0   
\end{bmatrix}
\end{equation}
It can be shown that modulation induces a shift of the left edge of the FBZ, thus the following relations hold:
\begin{equation}
\mu_B = -\pi (1-\nu_m ) \qquad \qquad \Omega_B =\frac{1}{2}(1-\nu_m )
\end{equation}
Moreover the expression for the upper and lower limits of the directional band gaps for backward propagating waves are:
\begin{align}
\Omega^{top}_B =&\frac{1}{2}\bigg[-\nu_m+\frac{1+\nu_m}{2}\sqrt{1+\frac{\alpha_m}{2}\frac{1-\nu_m}{1+\nu_m}}+\frac{1-\nu_m}{2}\sqrt{1+\frac{\alpha_m}{2}\frac{1+\nu_m}{1-\nu_m}}\bigg]\\
\Omega^{bot}_B =&\frac{1}{2}\bigg[-\nu_m+\frac{1+\nu_m}{2}\sqrt{1-\frac{\alpha_m}{2}\frac{1-\nu_m}{1+\nu_m}}+\frac{1-\nu_m}{2}\sqrt{1-\frac{\alpha_m}{2}\frac{1+\nu_m}{1-\nu_m}}\bigg]
\end{align}
while the corresponding linearized expressions write:
\begin{align}
\Omega^{top}_{B,lin}=&+\frac{\alpha_m}{8} +	\frac{1}{2}\Big(1-\nu_m\Big)\\
\Omega^{bot}_{B,lin}=&-\frac{\alpha_m}{8} +	\frac{1}{2}\Big(1-\nu_m\Big)
\end{align}
The linearized expression for the backward directional band gap width is given by:
\begin{equation}
\Delta\Omega_{B,lim}=\Omega^{top}_{B,lin}-\Omega^{bot}_{B,lin}=\frac{\alpha_m}{4}
\end{equation}
hence:
\begin{equation}
\Delta\Omega_{F,lim}=\Delta\Omega_{B,lim}=\frac{\alpha_m}{4}
\end{equation}
both backward and forward directional band gap widths have the same value, which depends on the modulation amplitude $\alpha_m$ only. We can also determine the minimum value of the modulation velocity parameter $\nu^{cr}_{m}$ such that, for $\alpha_m\neq0$, the beam shows two distinct directional band gaps, one associated to the backward propagating waves, the other one associated to the forward propagating waves. In order to obtain an expression for $\nu^{cr}_{m}$ we impose:
\begin{equation}
\Omega^{top}_{B,lim}=\Omega^{bot}_{F,lim}
\end{equation}
which, solved for $\nu_m$ gives:
\begin{equation}
\nu^{cr}_{m}=\frac{\alpha_m}{4}
\end{equation}
This results shows that, for smalls values of $\alpha_m$, the modulation velocity parameter $\nu_m$ determines a shift $\Omega_{shift}$ in the location of the band gaps that writes $\Omega_{shift}=\nu_m$ for forward propagating waves and $\Omega_{shift}=-\nu_m$
for backward propagating waves, thus breaking the mirror symmetry of the band diagram about the $\Omega$ axis.

Following the same procedure outlined in the case of longitudinal motion, we characterize backward and forward wave propagation in the case of transverse motion through approximate analytic expressions for the directional band gaps.
\begin{figure}[hbtp]
	\centering
	\begin{subfigure}[b]{0.48\textwidth}
		\includegraphics[width=\textwidth]{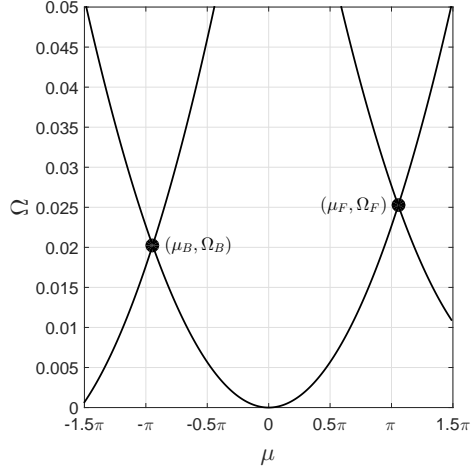}
		\caption{}
		\label{Fig:X}
	\end{subfigure}
	
	\begin{subfigure}[b]{0.48\textwidth}
		\includegraphics[width=\textwidth]{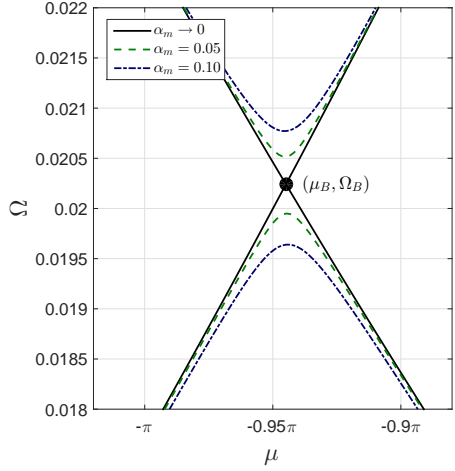}
		\caption{}
		\label{Fig:XX}
	\end{subfigure}	
		\centering
	\begin{subfigure}[b]{0.48\textwidth}
		\includegraphics[width=\textwidth]{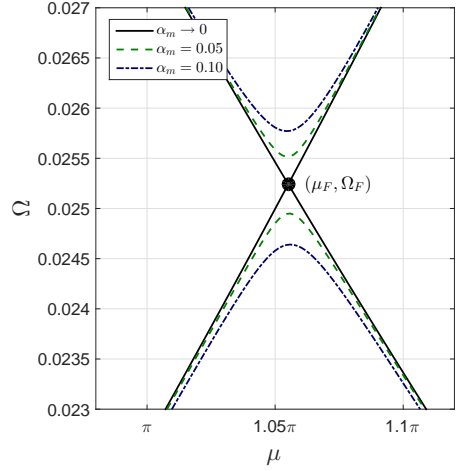}
		\caption{}
		\label{Fig:XXX}
	\end{subfigure}                   
	\caption{Band diagrams for transverse motion with $\nu_m=0.005$ and different values of $\alpha_m$. For $\alpha_m\to0$ (a), the first two dispersion branches intersect at $(\mu_B,\Omega_B)$ and $(\mu_F,\Omega_F)$ for backward and forward propagating waves, respectively. Comparison with the case in which $\alpha_m\neq0$ is given in (b) for backward propagating  waves and in (c) forward propagating waves.}
	\label{Fig:Analytical_Beam}
\end{figure}
We study the following QEP:
\begin{equation}\label{eq:beamQEP}
\begin{bmatrix} 
T_{11} & T_{12}  & T_{13}  \\ 
T_{21} & T_{22}  & T_{23}  \\ 
T_{31} & T_{32}  & T_{33} 
\end{bmatrix} 
\begin{bmatrix} 
\hat{w}_{-1}   \\ 
\hat{w}_{0}    \\ 
\hat{w}_{+1}   
\end{bmatrix} =
\begin{bmatrix} 
0  \\ 
0  \\ 
0   
\end{bmatrix}
\end{equation}
which we obtain from Eq.~\eqref{eq:QEP_T} by retaining the $0$-th and first order harmonics in Eq.~\eqref{eq:sol2}. The dimensionelss expressions for the coefficients in Eq.~\eqref{eq:beamQEP} write:
\begin{align}
T_{11} &= \Omega^2 - 2 \nu_m \Omega - \Big[\Big(\frac{\mu}{2 \pi}-1\Big)^4\chi^2 - \nu_m^2\Big] \\
T_{12} &=  -\Big[\frac{\mu}{2 \pi}\Big(\frac{\mu}{2 \pi}-1\Big)\Big]^2\frac{\alpha_m}{2} \\
T_{13} &=  0 \\
T_{21} &=  -\Big[\frac{\mu}{2 \pi}\Big(\frac{\mu}{2 \pi}-1\Big)\Big]^2\frac{\alpha_m}{2}  \\
T_{22} &=  \Omega^2 - \Big(\frac{\mu}{2 \pi}\Big)^4\chi^2 \\
T_{23} &=   -\Big[\frac{\mu}{2 \pi}\Big(\frac{\mu}{2 \pi}+1\Big)\Big]^2\frac{\alpha_m}{2} \\
T_{31} &=  0 \\
T_{32} &=    -\Big[\frac{\mu}{2 \pi}\Big(\frac{\mu}{2 \pi}+1\Big)\Big]^2\frac{\alpha_m}{2} \\
T_{33} &=  \Omega^2 + 2 \nu_m \Omega - \Big[\Big(\frac{\mu}{2 \pi}+1\Big)^4\chi^2 - \nu_m^2\Big]
\end{align}
in which we define $\chi=R_g \kappa_m$.
For forward wave propagation, Fig.~\ref{Fig:Analytical_Beam} shows that for $\nu_m=0.005$, the right edge of the FBZ between the first and the second dispersion branch is shifted due to modulation. Also, as as $\alpha_m\to0$, the band gap closes and the dispersion branches intersect at the point corresponding to the following dimensionless wavenumber and frequency:
\begin{equation}\label{eq:muFOmF_beam}
\mu_F = \pi \Big(1+\frac{\nu_m}{\chi} \Big) \qquad \qquad \Omega_F =\frac{1}{4}\Big(1+\frac{\nu_m}{\chi}\Big)^2\chi.
\end{equation}
in which $\mu_F\to\pi$ and $\Omega_F \to\chi/4$ as $\nu_m\to0$, hence  when no temporal modulation is applied. The limits of the directional band gaps write:
\begin{align}
\Omega^{top}_F =& \frac{1}{2}\bigg[\nu_m+\frac{1}{4}\bigg(1-\frac{\nu_m}{\chi}\bigg)^2\chi\sqrt{1+\frac{\alpha_m}{2}\bigg(\frac{1+\frac{\nu_m}{\chi}}{1-\frac{\nu_m}{\chi}}\bigg)^2}+\frac{1}{4}\bigg(1+\frac{\nu_m}{\chi}\bigg)^2\chi\sqrt{1+\frac{\alpha_m}{2}\bigg(\frac{1-\frac{\nu_m}{\chi}}{1+\frac{\nu_m}{\chi}}\bigg)^2}\bigg]\\
\Omega^{bot}_F =& \frac{1}{2}\bigg[\nu_m+\frac{1}{4}\bigg(1-\frac{\nu_m}{\chi}\bigg)^2\chi\sqrt{1-\frac{\alpha_m}{2}\bigg(\frac{1+\frac{\nu_m}{\chi}}{1-\frac{\nu_m}{\chi}}\bigg)^2}+\frac{1}{4}\bigg(1+\frac{\nu_m}{\chi}\bigg)^2\chi\sqrt{1-\frac{\alpha_m}{2}\bigg(\frac{1-\frac{\nu_m}{\chi}}{1+\frac{\nu_m}{\chi}}\bigg)^2}\bigg]
\end{align}
The radical in the previous expression can be linearized to get:
\begin{align}
\Omega^{top}_{F,lin}=&\frac{1}{2}\bigg[\nu_m + \frac{1}{2}\bigg(1+\frac{\alpha_m}{4}\bigg)\chi\bigg]\\
\Omega^{bot}_{F,lin}=&\frac{1}{2}\bigg[\nu_m + \frac{1}{2}\bigg(1-\frac{\alpha_m}{4}\bigg)\chi\bigg]
\end{align}
in which we also assume that $(\nu_m/\chi)^2<<1$. The linearized expression for the band gap width writes:
\begin{equation}
\Delta\Omega_{F,lim}=\Omega^{top}_{F,lin}-\Omega^{bot}_{F,lin}=\frac{\alpha_m}{8}\chi
\end{equation}.

For waves traveling backward, one can show that, for the first band gap with $\alpha_m\to0$, the modulation shifts the left edge of the FBZ leading to the following values:
\begin{equation}
\mu_B=-\pi\Big(1-\frac{\nu_m}{\chi}\Big) \qquad \qquad \Omega_B = \frac{1}{4}\Big(1-\frac{\nu_m}{\chi}\Big)^2\chi.
\end{equation}
For $\alpha_m<<1$ but different from zero, directional band gaps open and their limiting frequencies can be obtained as functions of the modulation parameters and written as follows:
\begin{align}
\Omega^{top}_B =& \frac{1}{2}\bigg[-\nu_m+\frac{1}{4}\bigg(1+\frac{\nu_m}{\chi}\bigg)^2\chi\sqrt{1+\frac{\alpha_m}{2}\bigg(\frac{1-\frac{\nu_m}{\chi}}{1+\frac{\nu_m}{\chi}}\bigg)^2}+\frac{1}{4}\bigg(1-\frac{\nu_m}{\chi}\bigg)^2\chi\sqrt{1+\frac{\alpha_m}{2}\bigg(\frac{1+\frac{\nu_m}{\chi}}{1-\frac{\nu_m}{\chi}}\bigg)^2}\bigg]\\
\Omega^{bot}_B =& \frac{1}{2}\bigg[-\nu_m+\frac{1}{4}\bigg(1+\frac{\nu_m}{\chi}\bigg)^2\chi\sqrt{1-\frac{\alpha_m}{2}\bigg(\frac{1-\frac{\nu_m}{\chi}}{1+\frac{\nu_m}{\chi}}\bigg)^2}+\frac{1}{4}\bigg(1-\frac{\nu_m}{\chi}\bigg)^2\chi\sqrt{1-\frac{\alpha_m}{2}\bigg(\frac{1+\frac{\nu_m}{\chi}}{1-\frac{\nu_m}{\chi}}\bigg)^2}\bigg]
\end{align}
Linearized expressions can be obtained and write:
\begin{align}
\Omega^{top}_{B,lin}=&\frac{1}{2}\bigg[-\nu_m + \frac{1}{2}\bigg(1+\frac{\alpha_m}{4}\bigg)\chi\bigg]\\
\Omega^{bot}_{B,lin}=&\frac{1}{2}\bigg[-\nu_m + \frac{1}{2}\bigg(1-\frac{\alpha_m}{4}\bigg)\chi\bigg]
\end{align}
therefore the band gap with has the following expression:
\begin{equation}
\Delta\Omega_{B,lim}=\Omega^{top}_{B,lin}-\Omega^{bot}_{B,lin}=\frac{\alpha_m}{8}\chi=\Delta\Omega_{F,lim}
\end{equation}
hence also for beams in transverse motion, the directional band gaps width has the same expression both for backward and forward propagating waves. Moreover the gap width does not depend on the modulation velocity parameter $\nu_m$, similarly to the case of longitudinal motion. Instead, the gap width depends on the modulation amplitude $\alpha_m$ and on the parameter $\chi$. We can compute the the critical modulation velocity to get:
\begin{equation}
\nu_m^{cr}=\frac{\alpha_m}{8}\chi
\end{equation}

\begin{figure}[hbtp]
	\centering
	\begin{subfigure}[b]{0.48\textwidth}
		\includegraphics[width=\textwidth]{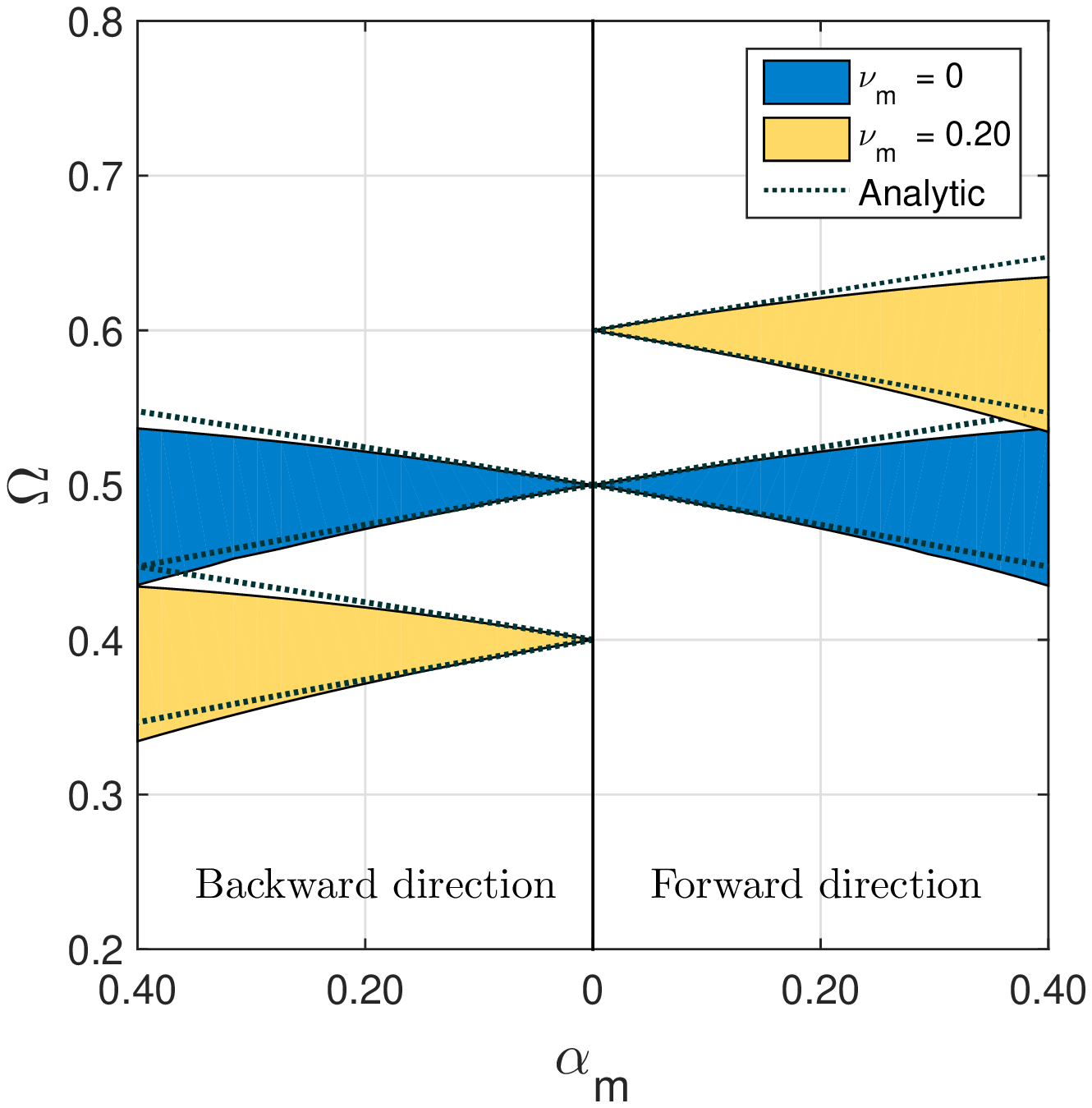}
		\caption{}
		\label{Fig:BGM_R_H_Zoom}
	\end{subfigure}         
	\begin{subfigure}[b]{0.48\textwidth}
		\includegraphics[width=\textwidth]{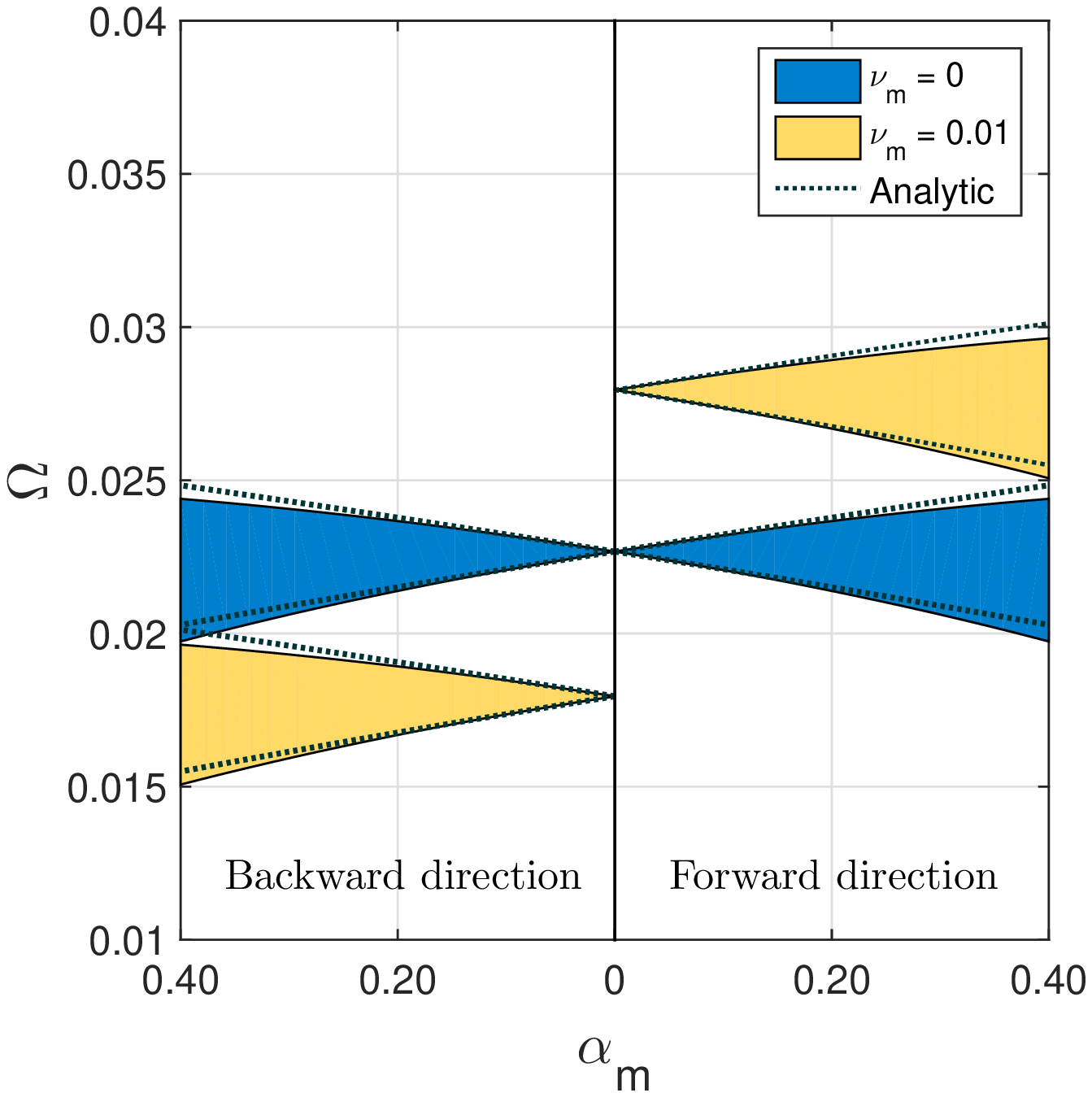}
		\caption{}
		\label{Fig:BGM_B_H_Zoom}
	\end{subfigure}               
	\caption{Band gap maps for beam in longitudinal (a) and transverse (b) motion. The gap amplitude is plotted as a function of the increasing modulation amplitude parameter $\alpha_m$ for the two cases $\nu_m=0$ and $\nu_m\neq0$, \emph{i.e.} with zero and nonzero modulation velocity parameter, respectively. The colored regions represent the band gap evolution as computed solving the full QEP, while the approximate analytic expression, valid for $\alpha_m<<1$, are represented by the dashed lines.}
	\label{Fig:BGM}
\end{figure}

We assess the validity of the approximate relations discussed above by
comparing the band gaps given by the full solution of the QEP with $N=3$ and by the analytic expression valid for $\alpha_m<<1$. Results are presented in Fig.\ref{Fig:BGM} for both longitudinal and transverse motion. We track the gap amplitude as a function of $\alpha_m$ for the two cases $\nu_m=0$ and $\nu_m\neq0$, \emph{i.e.} with zero and nonzero modulation velocity parameter, respectively. For $\nu_m=0$ the system shows complete band gaps originating from the same value of frequency $\Omega$ as $\alpha_m\to0$, thus symmetry with respect to backward and forward propagation holds. On the contrary, for $\nu_m\neq0$ symmetry is broken, we have directional band gaps that originate from different values of frequency $\Omega$ as $\alpha_m\to0$. We observe that the analytic formulae for the band gaps not only represent an excellent approximation for $\alpha_m<<1$, but they also appear to give accurate estimates when $\alpha_m$ is not small compared to unity.
\subsection{Numerical simulations}
\begin{figure}[hbtp]
	\centering
	\begin{subfigure}[b]{0.48\textwidth}
		\includegraphics[width=\textwidth]{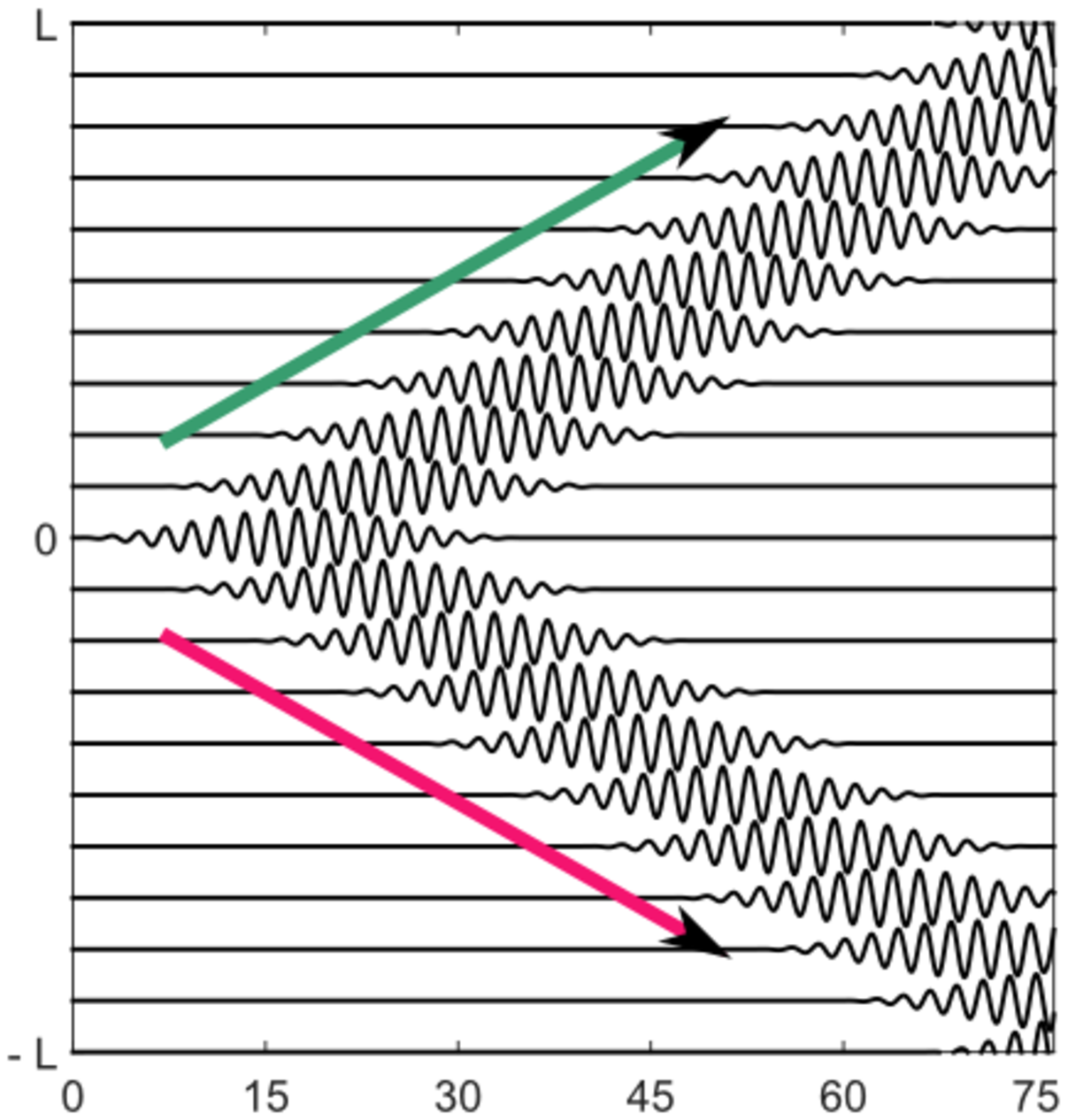}
		\caption{}
		\label{Fig:Waterfall_NoMod}
	\end{subfigure}         
	\begin{subfigure}[b]{0.48\textwidth}
		\includegraphics[width=\textwidth]{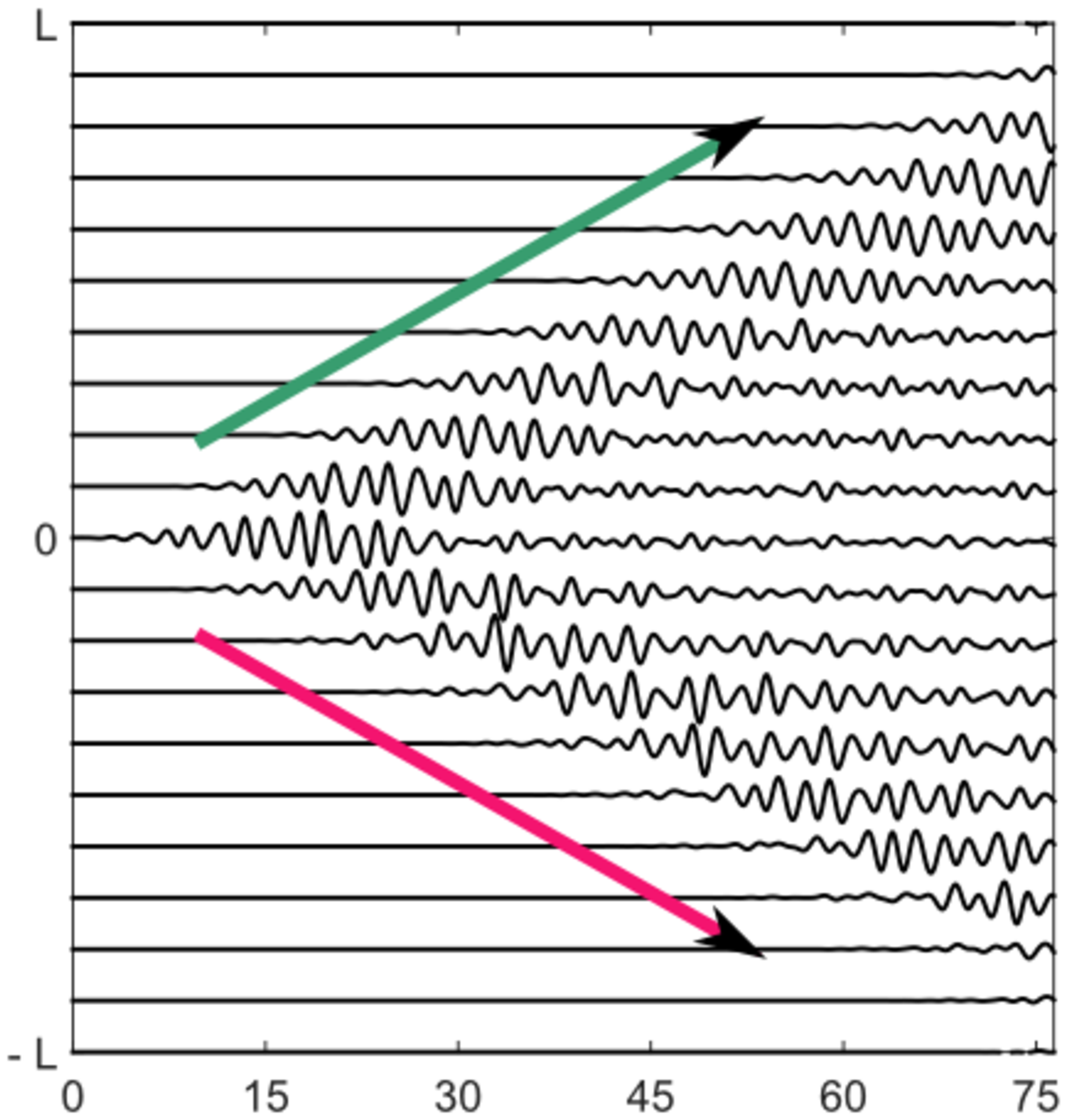}
		\caption{}
		\label{Fig:Waterfall_Mod_PassBand_Rod}
	\end{subfigure}               
	\begin{subfigure}[b]{0.48\textwidth}
		\includegraphics[width=\textwidth]{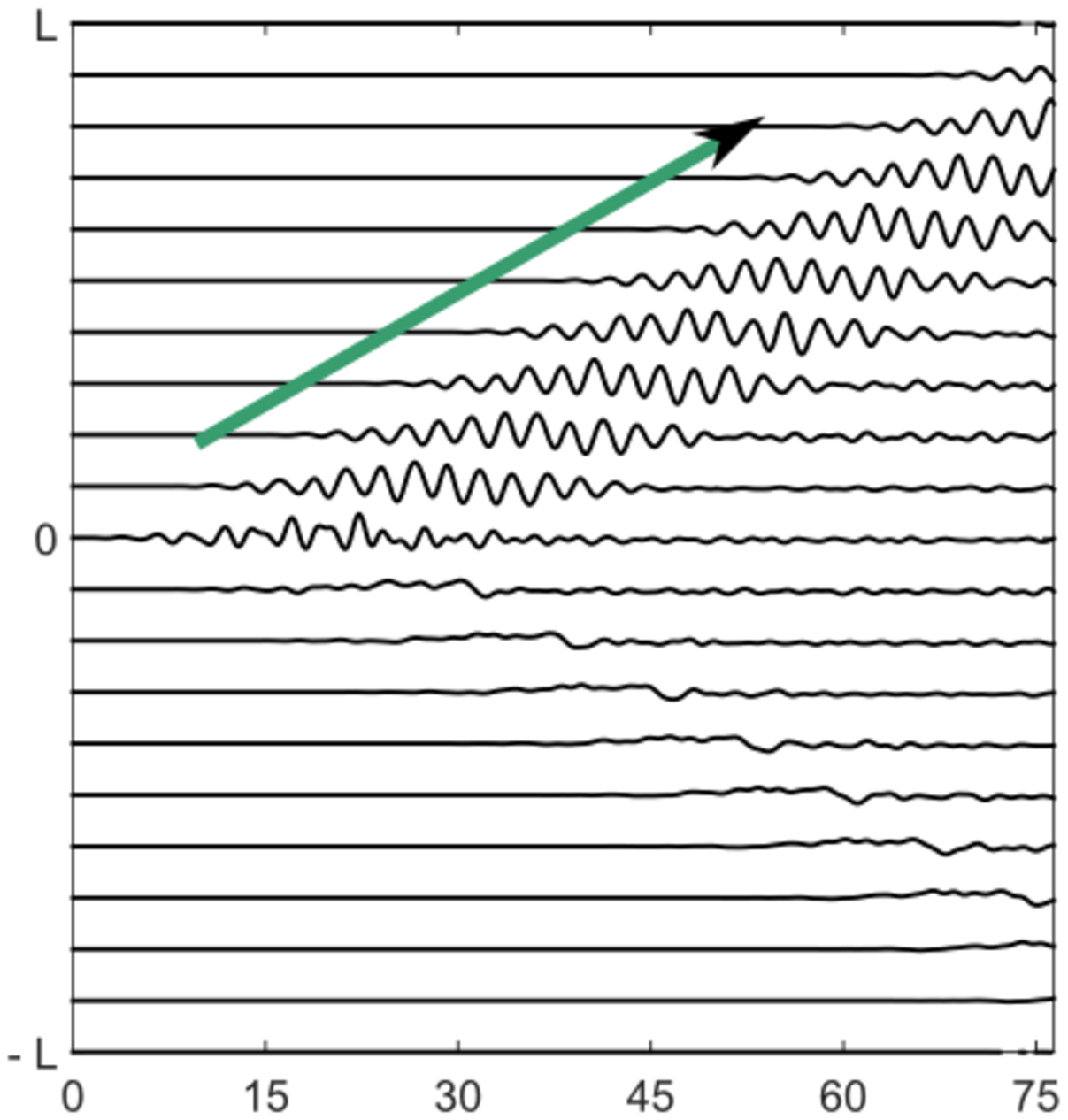}
		\caption{}
		\label{Fig:Waterfall_Mod_NoBackward_Rod}
	\end{subfigure}
		\begin{subfigure}[b]{0.48\textwidth}
		\includegraphics[width=\textwidth]{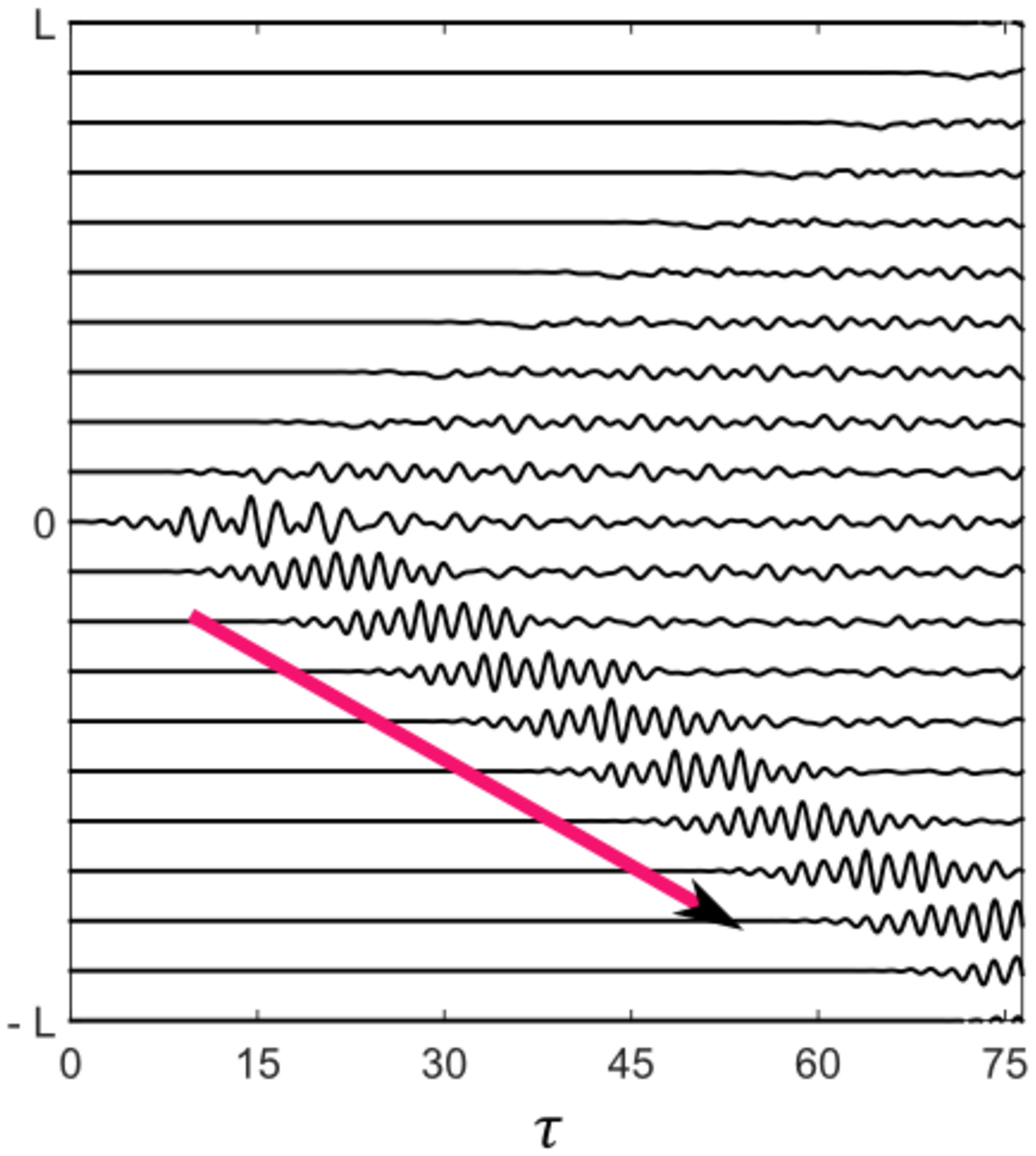}
		\caption{}
		\label{Fig:Waterfall_Mod_NoForward_Rod}
	\end{subfigure}
	\caption{Waterfall plots representing the transient response of a $2L$ long beam in longitudinal motion to narrow band external excitation loaded at its mid span. In (a), we have both forward and backward waves in a non-modulated beam excited around $\Omega_{ext}=0.485$; in (b),we impose  modulation with $\alpha_m=0.40$ and $\nu_m=0.20$ and same excitation, waves in both directions propagate; in (c) the same modulated beam is excited with at a frequency centered at $\Omega_{ext}=0.3844$, which falls within a directional band gap, thus forward-propagating waves are supported by the structure; in (d) the signal of the load is centered at $\Omega_{ext}=0.5844$, therefore backward-propagating waves only are allowed.}
	\label{Fig:Waterfall_Rod}
\end{figure}

\begin{figure}[hbtp]
	\centering
	\begin{subfigure}[b]{0.48\textwidth}
		\includegraphics[width=\textwidth]{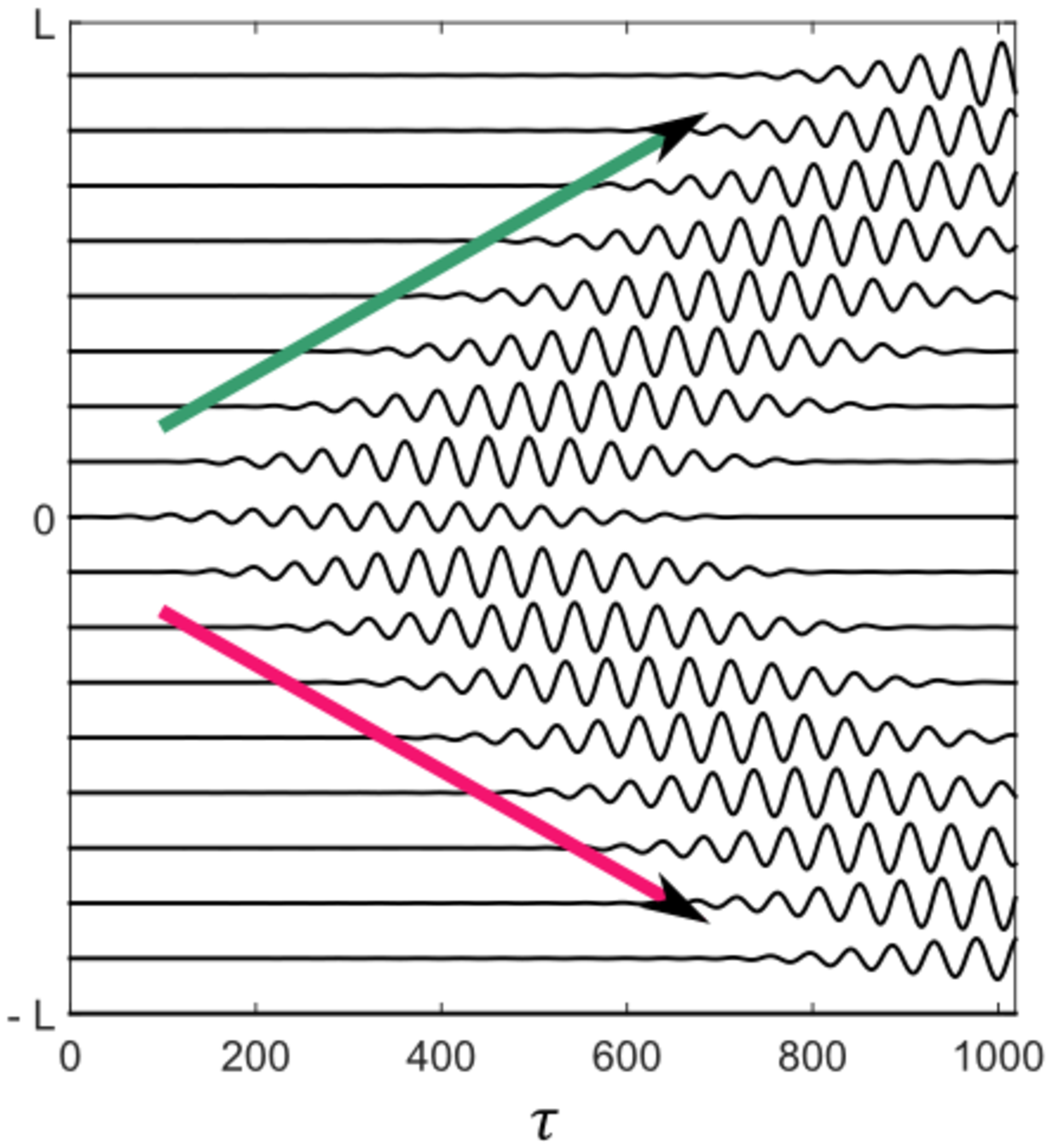} 
		\caption{}
		\label{Fig:Waterfall_NoMod_Beam}
	\end{subfigure}         
	\begin{subfigure}[b]{0.48\textwidth}
		\includegraphics[width=\textwidth]{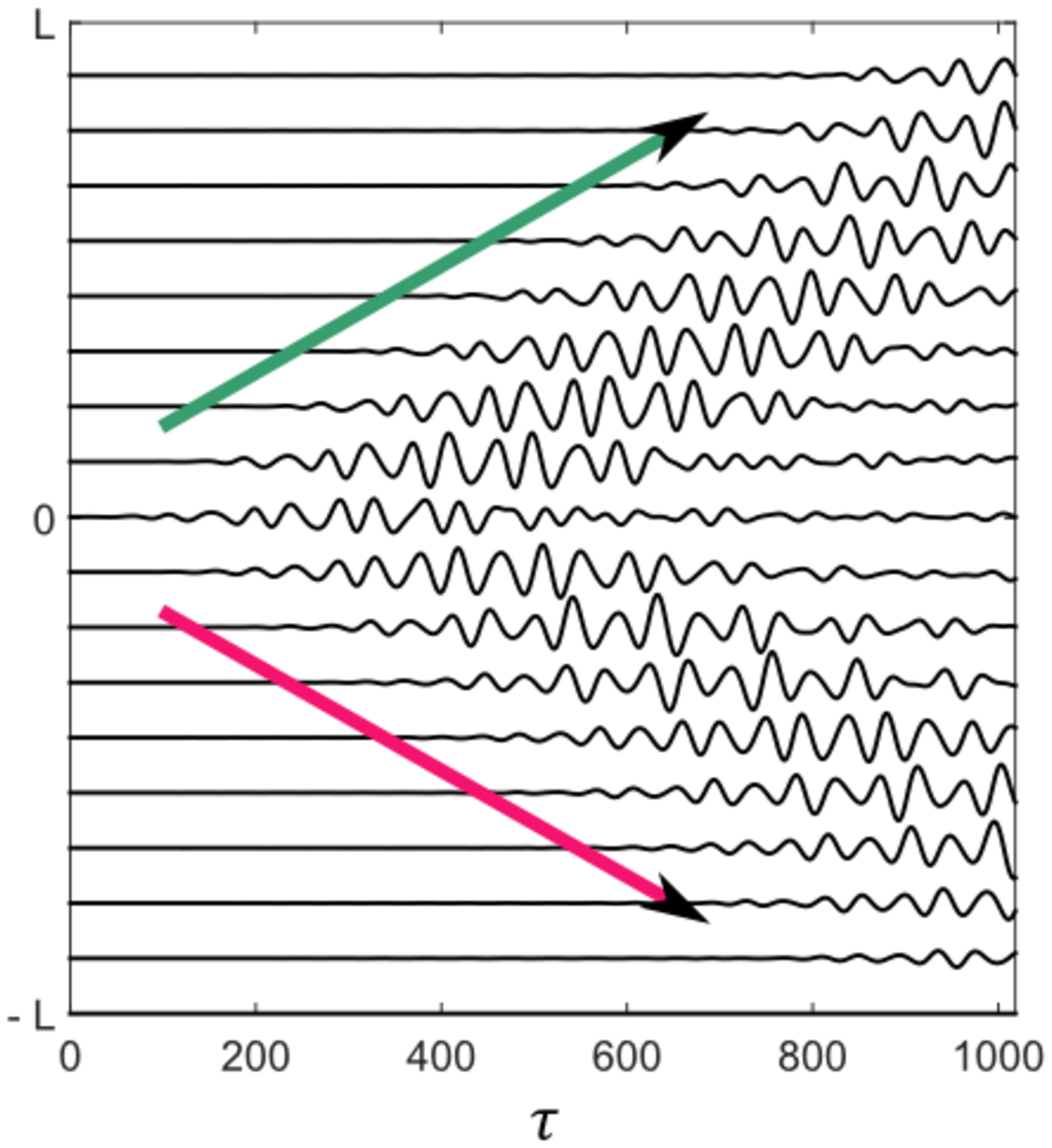}
		\caption{}
		\label{Fig:Waterfall_Mod_PassBand_Beam}
	\end{subfigure}               
	\begin{subfigure}[b]{0.48\textwidth}
		\includegraphics[width=\textwidth]{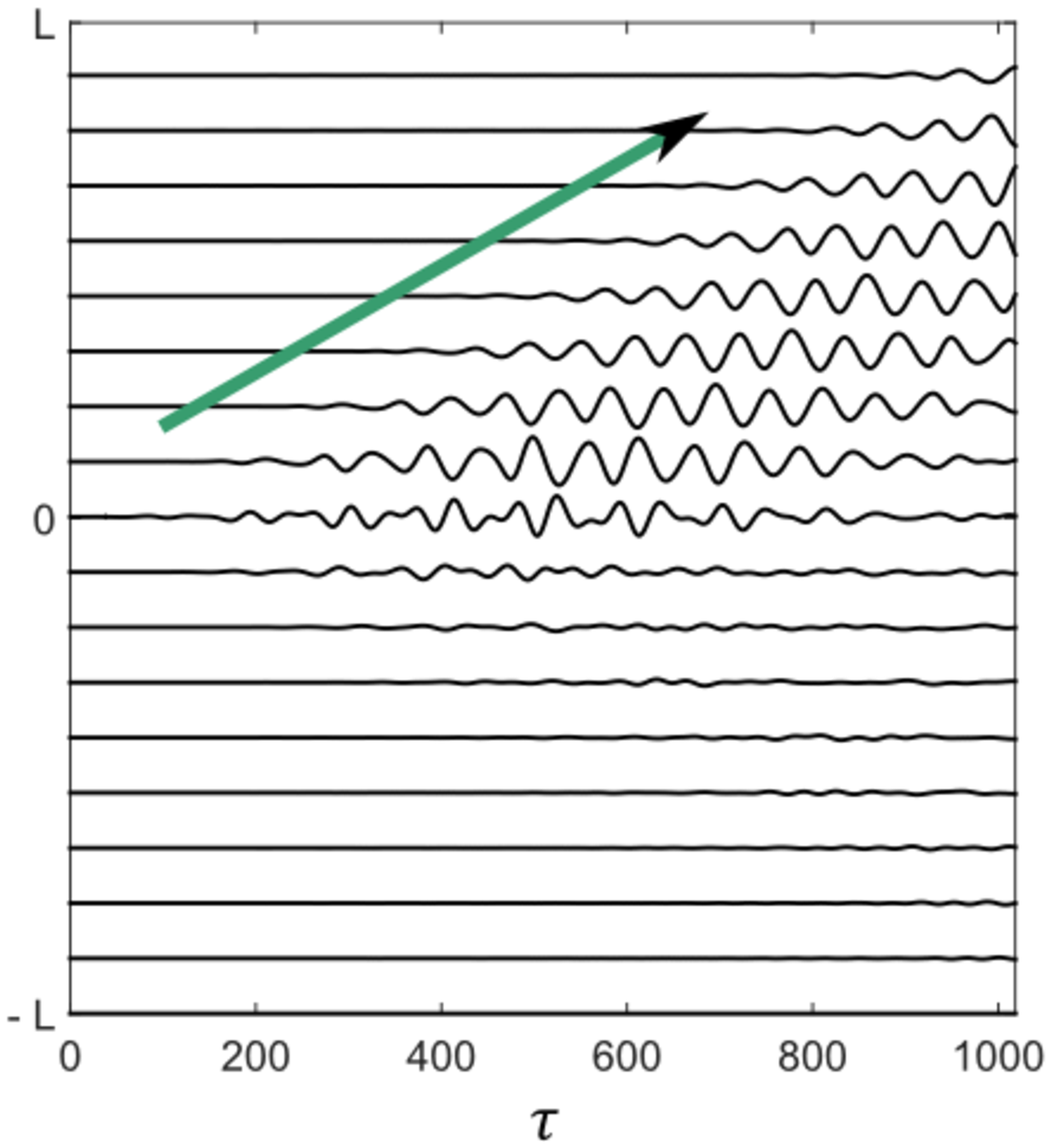}
		\caption{}
		\label{Fig:Waterfall_Mod_NoBackward_Beam}
	\end{subfigure}
		\begin{subfigure}[b]{0.48\textwidth}
		\includegraphics[width=\textwidth]{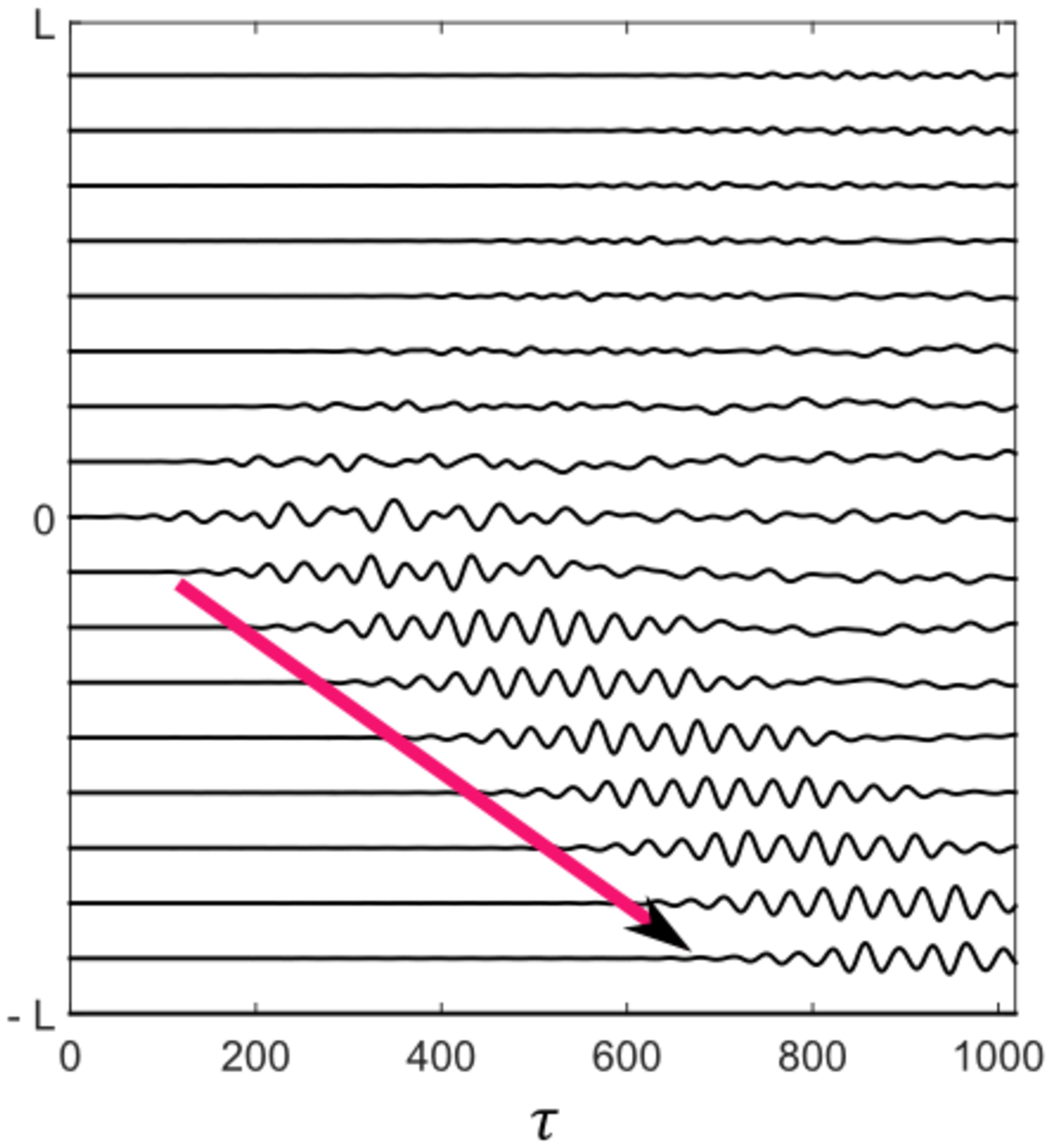}
		\caption{}
		\label{Fig:Waterfall_Mod_NoForward_Beam}
	\end{subfigure}
	\caption{Waterfall plots representing the transient response of a $2L$ long beam in transverse motion to narrow band external excitation loaded at its mid span, with $\chi=0.0144$. Reciprocal behavior is shown in (a) when no modulation is applied and the excitation frequency content is centered at $\Omega_{ext}=0.0225$; in (b), the same excitation is applied to a modulated beam with $\alpha_m=0.40$ and $\nu_m=0.01$ but the structure behaves still reciprocally; in (c) the same modulated beam is excited with a load having frequency centered at $\Omega_{ext}=0.0175$, which falls within a directional band gap, thus forward-propagating waves are supported by the structure; in (d) the signal of the load is centered at $\Omega_{ext}=0.0275$, therefore backward-propagating waves only are allowed.}
	\label{Fig:Waterfall_Beam}
\end{figure}

We perform numerical simulations to verify the non-reciprocal behavior of the beams when an external excitation is imposed. Specifically, we compute the transient response of the structure to narrow-band excitation and analyze the displacement field. The computation are performed by using the commercial finite element code COMSOL Multiphysics. We consider a $2L$ long beam, with $L=70\lambda_m$ and a total of $140$ unit cells. In our model, each unit cell is discretized by 20 quadrilateral elements, therefore we consider $2800$ elements for the whole structure. In this section we assume $\chi=0.0144$. The excitation is imposed halfway through the length of the structures at $x=0$ as a point load, so that both backward and forward propagating waves are excited. The load acts along the $x$ direction when we target longitudinal motion, while it acts normally to the $x$ direction for transverse motion.  We introduce the dimensionless time $\tau$ as:
\begin{equation}
\tau=\frac{c_0 t}{\lambda_m}
\end{equation}
which represents the number of unit cells of length $\lambda_m$ that a longitudinal wave, traveling at the speed $c_0$ in a uniform beam, covers in the time $t$.
\begin{figure}
  \centering
    \includegraphics[width=0.48\textwidth]{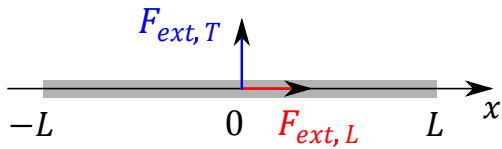}
      \caption{Schemtic of a $2L$ long beam loaded in its mid span. Longitudinal and transverse motion are excited by the horizontal force $F_{ext,L}$ and the transverse force $F_{ext,T}$, respectively.}
      \label{fig:BeamSchematic_ExternalExcitation}
\end{figure}
Results are presented in Fig.~\ref{Fig:Waterfall_Rod} and Fig.~\ref{Fig:Waterfall_Beam} in the form of waterfall plots, that allow us to track how the waves propagate throughout the system. We compare a non-modulated beam, in which we consider $\alpha_m=0$, $\nu_m=0$, with a spatiotemporal modulated beam with $\alpha_m=0.40$ and $\nu_m=0.20$ for longitudinal motion and with $\alpha_m=0.40$ and $\nu_m=0.01$ for transverse motion. The results can be better interpreted when compared to the band diagrams of the corresponding system in Fig.~\ref{Fig:BD_R_H_01_BG_noNegFreq1dot5pi} and Fig.~\ref{Fig:BD_B_H_01_BG_noNegFreq1dot5pi_Leg_Om0dot05} for the non-modulated beam and Fig.~\ref{Fig:BD_R_H_04_BG_noNegFreq1dot5pi} and Fig.~\ref{Fig:BD_B_H_04_BG_noNegFreq1dot5pi_Leg_Om0dot05} for the modulated beams, respectively. When no modulation is applied, the system is perfectly reciprocal and waves propagate symmetrically in the backward and forward directions, as shown in Fig.~\ref{Fig:Waterfall_NoMod} and Fig.~\ref{Fig:Waterfall_NoMod_Beam}, with excitation centered at $\Omega=0.485$ and $\Omega=0.0225$ for longitudinal and transverse motion, respectively. When the same excitation is applied but the properties of the beam are modulated, the systems still behaves in a reciprocal fashion since the excitation is centered in a pass band, as shown in band diagram in Fig.~\ref{Fig:BD_R_H_04_BG_noNegFreq1dot5pi}. The waterfall plots for this case are shown in Fig.~\ref{Fig:Waterfall_Mod_PassBand_Rod} and Fig.~\ref{Fig:Waterfall_Mod_PassBand_Beam} for longitudinal and transverse motion, respectively. Conversely, strong one-directional wave propagation is achieved when the excitation is centered in a directional band gap. We recognize forward-only propagation in Fig.~\ref{Fig:Waterfall_Mod_NoBackward_Rod} and Fig.~\ref{Fig:Waterfall_Mod_NoBackward_Beam}, being the modulated structure excited with a narrow band signal centered at $\Omega=0.3844$ and $\Omega=0.0175$, for longitudinal and transverse motion, respectively. Instead, backward-only propagation can be observed in Fig.~\ref{Fig:Waterfall_Mod_NoForward_Rod} and Fig.~\ref{Fig:Waterfall_Mod_NoForward_Beam} for longitudinal and transverse motion when the excitation is centered at $\Omega=0.5844$ and $\Omega=0.0275$, respectively.

Another numerical approach to characterize the dispersion properties of a beam consists in exciting the structure and computing the two-dimensional Fourier transform (2DFT) of the recorded displacement fields $u(x,t)$ and $w(x,t)$:
\begin{align}\label{eq:2DFTU}
U(\kappa,\omega)=&\int_{-\infty}^{\infty}\int_{-\infty}^{\infty} u(x,t) e^{-i(\omega t - \kappa x)} dx \, dt\\
\label{eq:2DFTW}
W(\kappa,\omega)=&\int_{-\infty}^{\infty}\int_{-\infty}^{\infty} w(x,t) e^{-i(\omega t - \kappa x)} dx \, dt 
\end{align}
The transformation allows one to describe the response of the beam in the wavenumber/frequency domain, effectively obtaining a band diagram for the structure. This technique is successfully applied in experimental validations as well~\cite{Airoldi2011}. We compare the band diagrams given by plotting the magnitude $|U(\kappa,\omega)|$
and $|W(\kappa,\omega)|$ of the complex quantities defined by Eq.~\eqref{eq:2DFTU} and~\eqref{eq:2DFTW}, in the form of contour lines, with the band diagrams that are given by the method discussed in the previous sections.

\begin{figure}[hbtp]
	\centering
	\begin{subfigure}[b]{0.48\textwidth}
		\includegraphics[width=\textwidth]{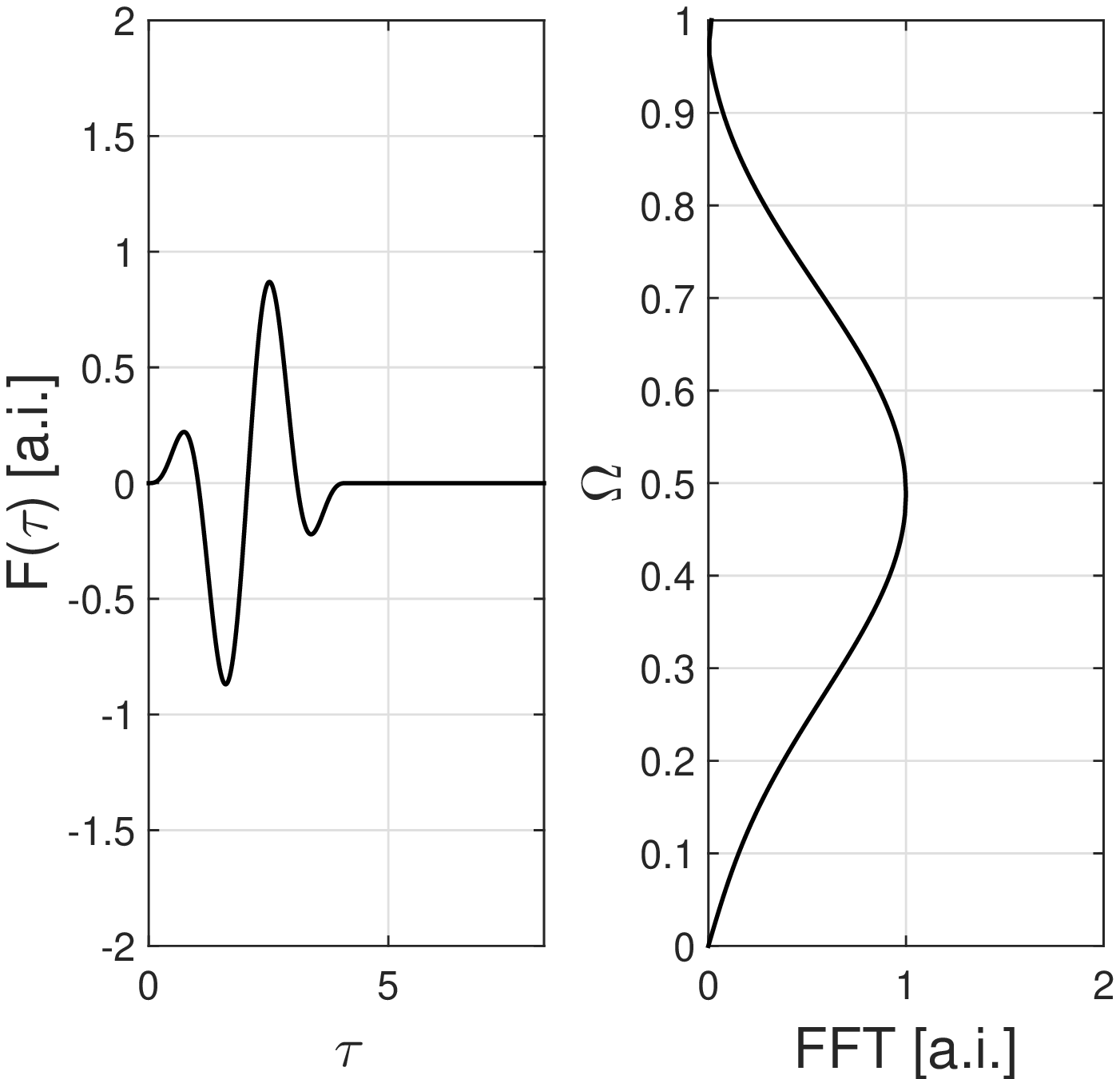}
		\caption{}
		\label{Fig:Signal_2DFT_rod}
	\end{subfigure}
	\begin{subfigure}[b]{0.48\textwidth}
		\includegraphics[width=\textwidth]{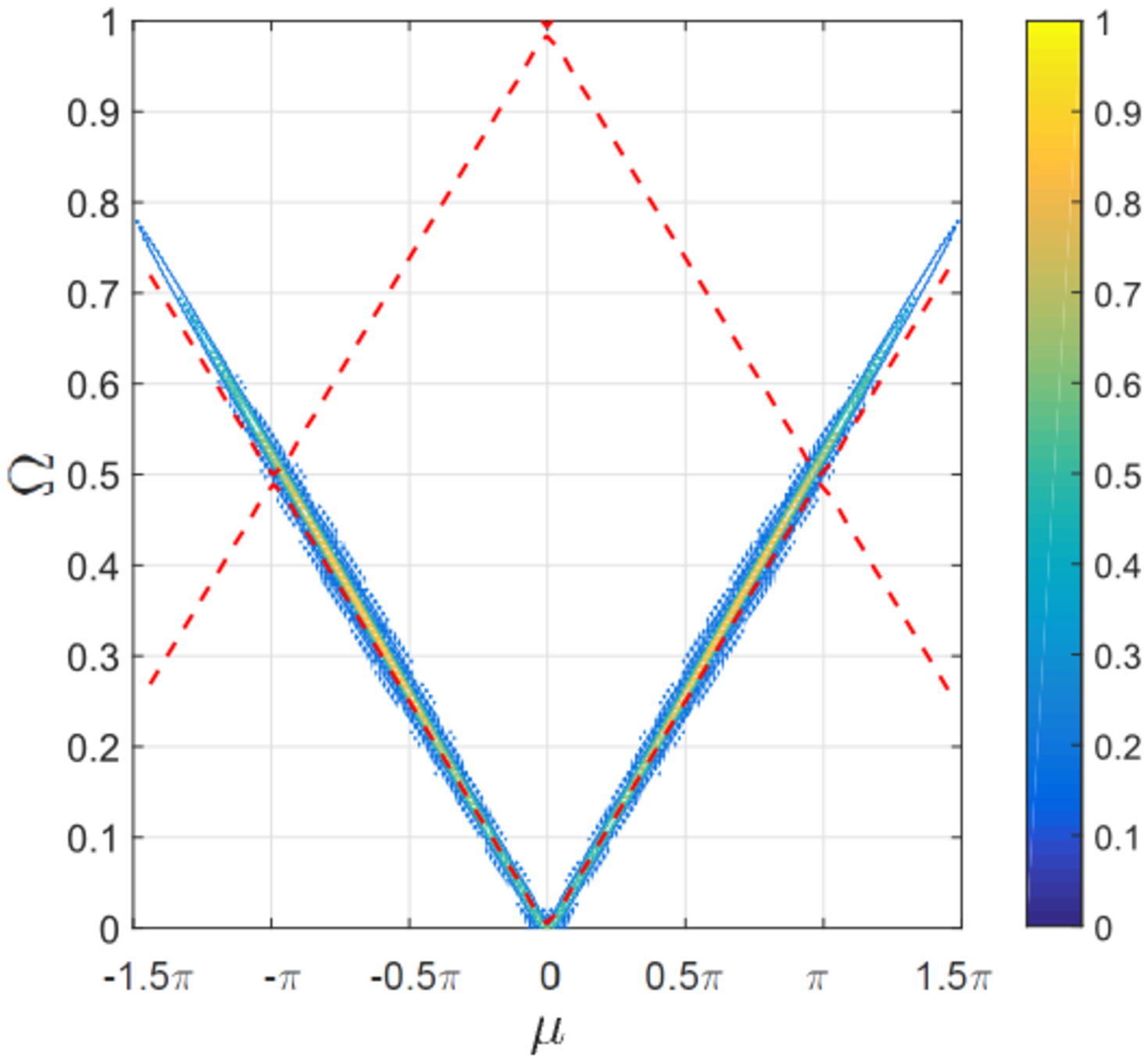}
		\caption{}
		\label{Fig:2DFT_EPWEM_rod_legend_NoMod_Dashed}
	\end{subfigure}
	
		\centering
	\begin{subfigure}[b]{0.48\textwidth}
		\includegraphics[width=\textwidth]{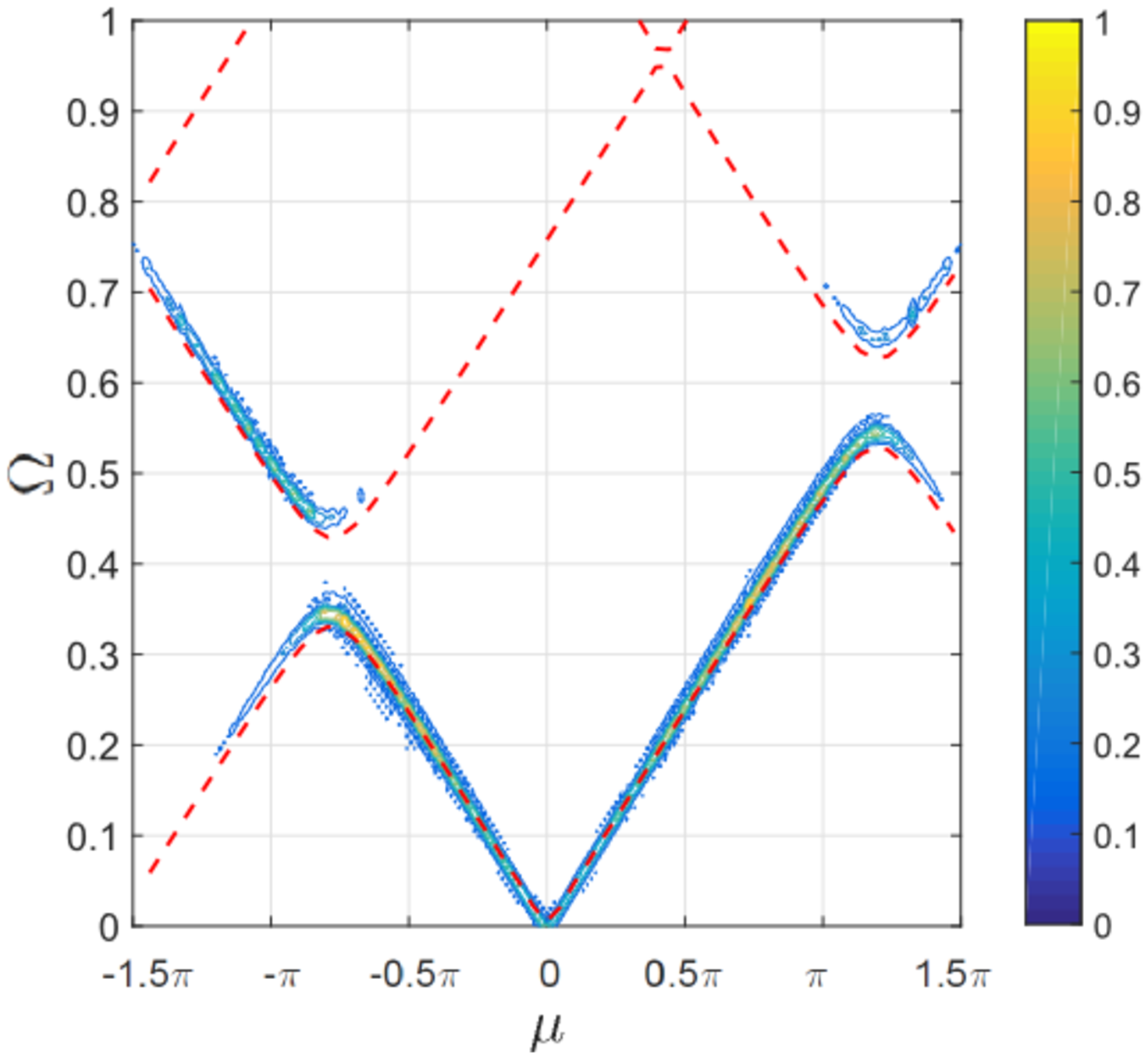}
		\caption{}
		\label{Fig:2DFT_EPWEM_rod_legend_Dashed}
	\end{subfigure}                   
	\caption{For longitudinal motion of $2L$ long beam excited at its mid span, comparison between the band diagram obtained solving the QEP (dashed lines) and through the normalized magnitude $|U(\kappa,\omega)|$ of the 2DFT of the displacement field (contour lines): (a) excitation signal and its frequency spectrum; (b) band diagram for uniform beam ($\alpha_m=0$ and $\nu_m=0$); (c) band diagram for modulated beam ($\alpha_m=0.40$ and $\nu_m=0.20$).}
	\label{Fig:2DFT_rod}
\end{figure}

\begin{figure}[hbtp]
	\centering
	\begin{subfigure}[b]{0.48\textwidth}
		\includegraphics[width=\textwidth]{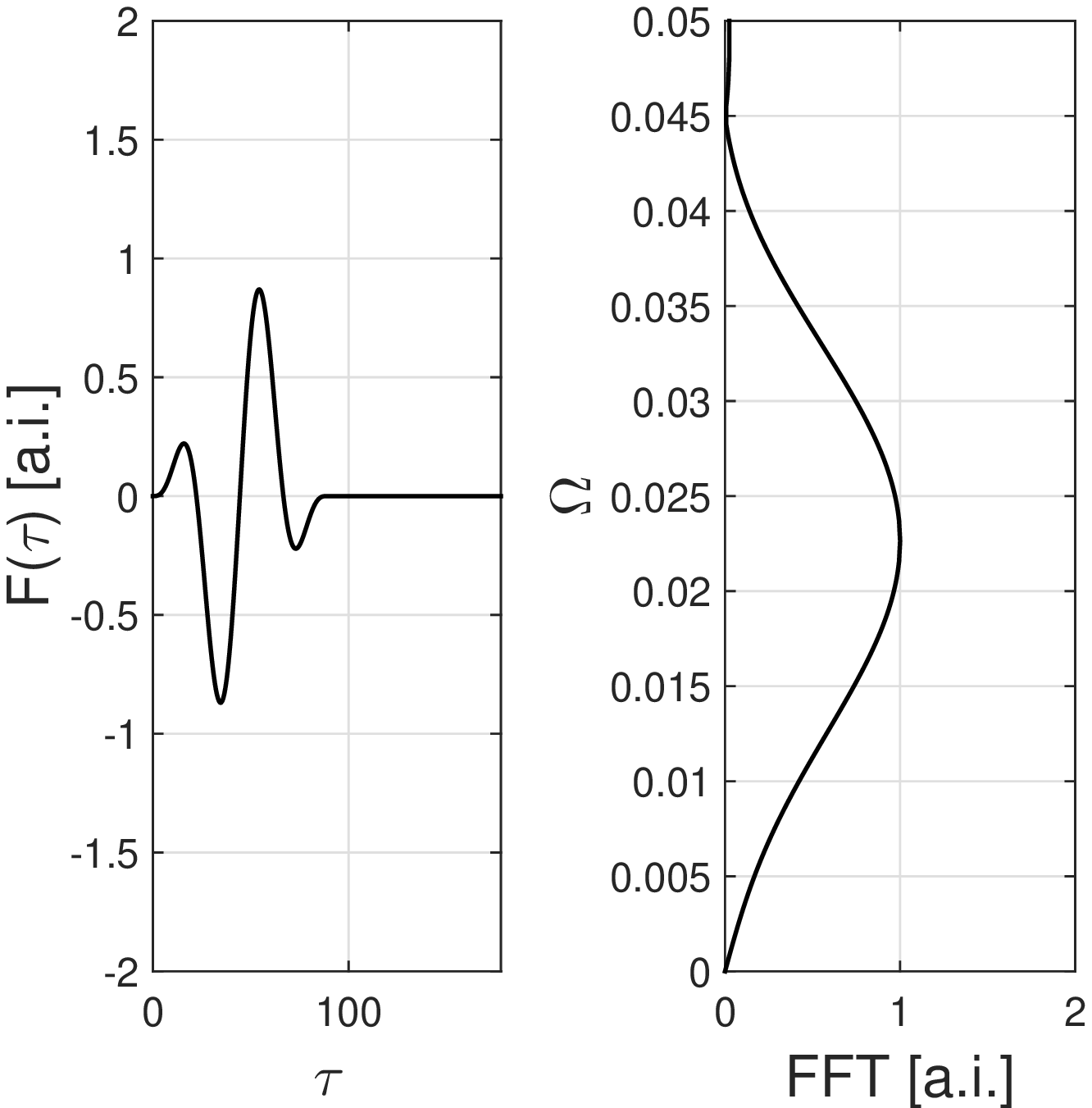}
		\caption{}
		\label{Fig:Signal_2DFT_beam}
	\end{subfigure}
	\begin{subfigure}[b]{0.48\textwidth}
		\includegraphics[width=\textwidth]{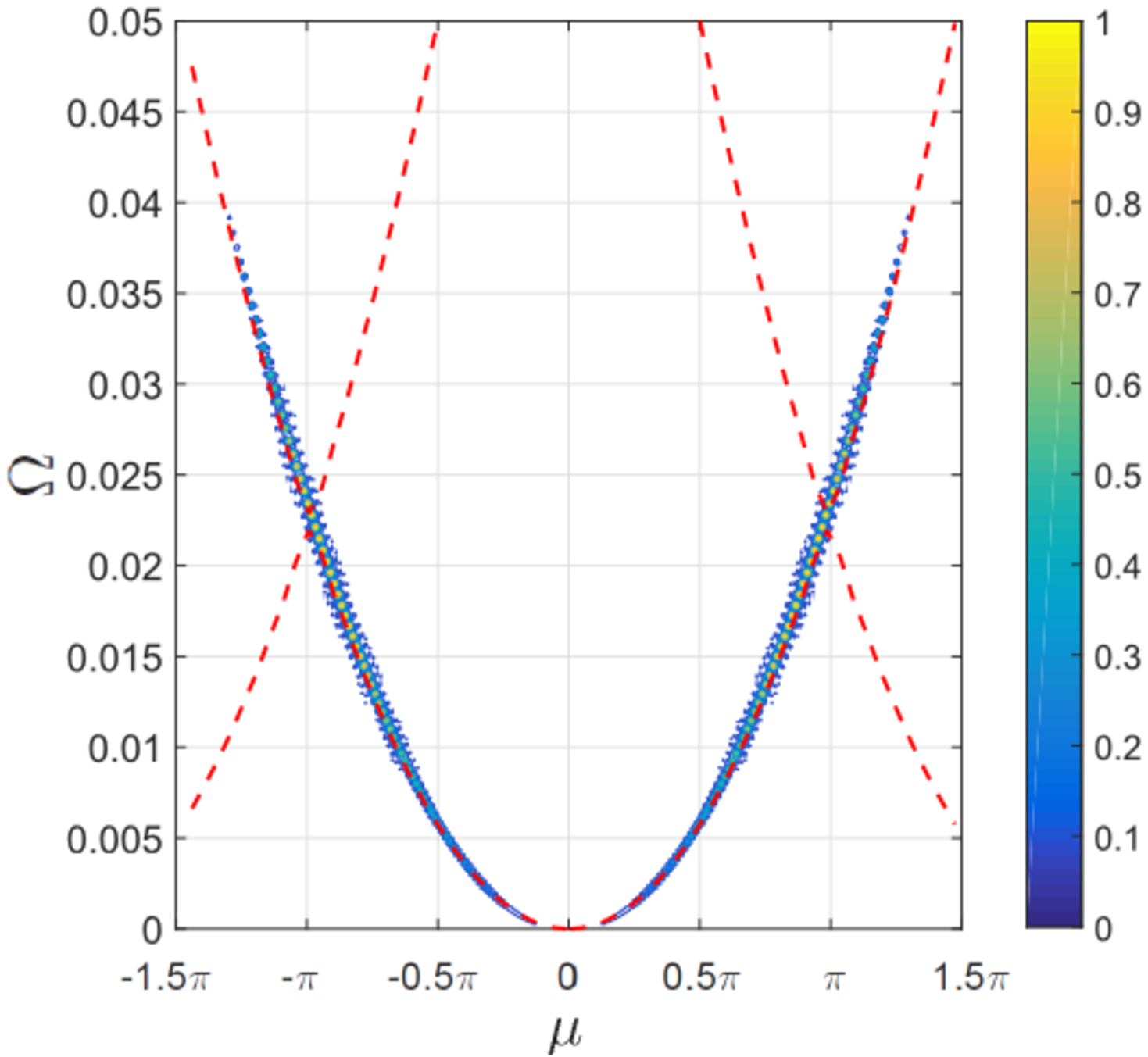}
		\caption{}
		\label{Fig:2DFT_EPWEM_beam_legend_NoMod_Dashed}
	\end{subfigure}
	
		\centering
	\begin{subfigure}[b]{0.48\textwidth}
		\includegraphics[width=\textwidth]{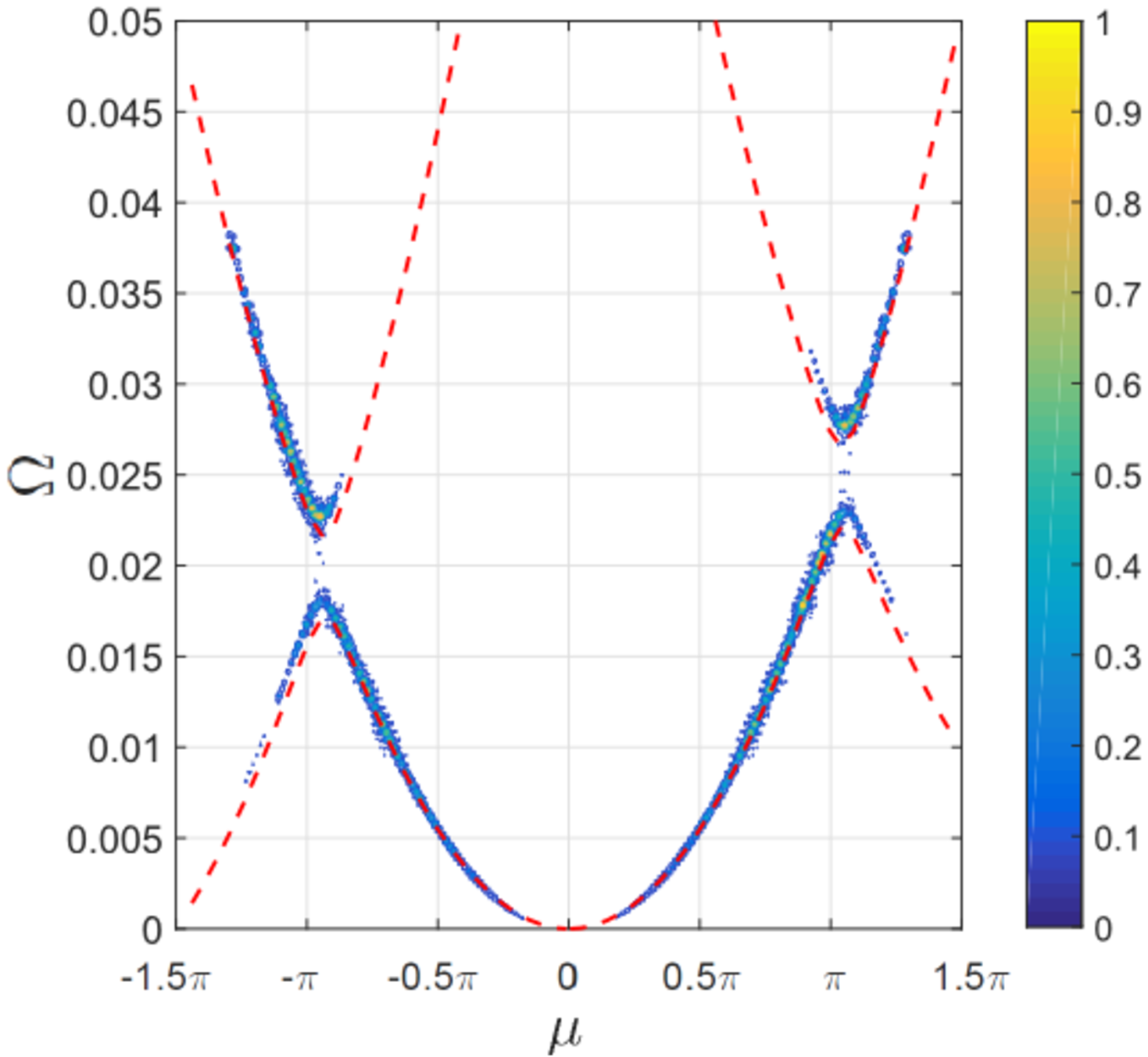}
		\caption{}
		\label{Fig:2DFT_EPWEM_beam_Nolegend_Dashes}
	\end{subfigure}            
	\caption{For transverse motion of $2L$ long beam excited at its mid span, comparison between the band diagram obtained solving the QEP (dashed lines) and through the normalized magnitude $|W(\kappa,\omega)|$ of the 2DFT of the displacement field (contour lines): (a) excitation signal and its frequency spectrum; (b) band diagram for uniform beam ($\alpha_m=0$ and $\nu_m=0$); (c) band diagram for modulated beam ($\alpha_m=0.40$ and $\nu_m=0.005$).}
	\label{Fig:2DFT_BEAM}
\end{figure}

The results are presented in Fig.~\ref{Fig:2DFT_EPWEM_rod_legend_NoMod_Dashed} and Fig.~\ref{Fig:2DFT_EPWEM_rod_legend_Dashed} for longitudinal motion of respectively a uniform and modulated beam with $\alpha_m=0.40$ and $\nu_m=0.20$. Similarly, Fig.~\ref{Fig:2DFT_EPWEM_beam_legend_NoMod_Dashed} and  Fig.~\ref{Fig:2DFT_EPWEM_beam_Nolegend_Dashes} show the results for transverse motion of a uniform and modulated beam with $\alpha_m=0.40$ and $\nu_m=0.005$, respectively. When longitudinal waves are concerned, the structure is excited with a load characterized by a broadband signal centered at $\Omega=0.485$ and acting along the $x$ direction as shown in Fig.~\ref{Fig:Signal_2DFT_rod}, while for transverse waves a vertical load centered at $\Omega=0.0225$ is used. The broadband signal is now required in order to excite the structure within the frequency range of interest, which in our case has to be wide enough to obtain the band diagram. We recognize that the contour lines used to plot both $|U(\kappa,\omega)|$ and $|W(\kappa,\omega)|$ overlap with the dispersion branches given by the proposed method, thus verifying the validity of the methodology for the calculation of the dispersion diagrams of spatiotemporal structures. For comparison, Fig.~\ref{Fig:2DFT_rod} and~\ref{Fig:2DFT_BEAM} show also the case in which the modulation is spatial only, hence for $\nu_m=0$.
\section{Conclusions}\label{ConclusionsSection}
In summary, we characterize dispersion properties of spatiotemporal periodic beams in longitudinal and transverse motion. By employing a solution in the Floquet form with space and time harmonics, we are able to compute the dispersion diagrams for such structures by solving a quadratic eigenvalue problem. The analysis of the dispersion diagrams allows us to describe the unique features that spatiotemporal modulation induces on the wave propagation properties of the structure, such as symmetry breaking of the dispersion relation and the relative dispersion diagrams. Specifically, we identify the signature of one-way propagation as directional band gaps. Such band gaps clearly describe in which frequency ranges forward-propagating or backward-propagating waves only are supported by the structure. The key finding of the study allow us to analytically relate the  modulation parameters to the position and width of the directional band diagrams, and also compute the minimum modulation speed required in order to have a fully non-reciprocal wave propagation for harmonic modulation. Finally, we verify our prediction with numerical simulations.

Nonreciprocal systems are the object of an emerging field of study and promise to deeply impact the way we control wave propagation, enriching the design space for technological applications in acoustics, phononics and photonics, to name a few. For this reason, when spatiotemporal periodic structures are used to achieve one-way prorogation, is it important to be  able to properly describe and characterize their non-reciprocal behavior, correctly predicting the influence of the modulation parameters. The proposed methodology was exposed in the context of elastic waves propagating in an elastic solid, but it can be easily applicable in any wave propagation problem which involves a periodic variation of the medium characteristics in both space and time.

\section{Acknowledgments}
\label{sec:Acknowledgments}
\emph{This work is conducted under the support of the Army Research Office, under grant No. W911NF1210460 monitored by Dr. David Stepp, whose support is greatly appreciated.}
\newpage
\bibliographystyle{unsrt}
\clearpage
\bibliography{papers}
\end{document}